\documentclass[journal]{IEEEtran}

\ifCLASSINFOpdf

\else

\fi

\usepackage{subfigure}
\usepackage{xcolor}
\usepackage{easyReview}
\usepackage[ruled]{algorithm2e} 
\usepackage{algorithmicx}
\usepackage{multirow} 
\usepackage{multicol} 

\usepackage[colorlinks,            
linkcolor=black,           
anchorcolor=black,            
citecolor=black]{hyperref}

\usepackage{lineno,hyperref}
\usepackage{color,soul}
\usepackage{bm}
\usepackage{amsthm,amsmath,amssymb}
\usepackage{mathrsfs}
\usepackage{overpic}
\usepackage{multirow}
\usepackage{booktabs}       % professional-quality tables
\usepackage{makecell}
\usepackage{array}
\usepackage{paralist}
\usepackage{subfigure}
\usepackage{graphicx}
\usepackage{pifont}
\usepackage{arydshln}
\modulolinenumbers[5]
\newcommand{\PreserveBackslash}[1]{\let\temp=\\#1\let\\=\temp}
\newcolumntype{C}[1]{>{\PreserveBackslash\centering}p{#1}}
\newcolumntype{R}[1]{>{\PreserveBackslash\raggedleft}p{#1}}
\newcolumntype{L}[1]{>{\PreserveBackslash\raggedright}p{#1}}
\usepackage{stfloats}

\def\revs1#1{{\color{blue}{\it{#1}}}}

\setreviewsoff

\begin{document}

\title{Arbitrary-shape Scene Text Detection via  Visual-Relational Rectification and Contour Approximation}

{
\author{Chengpei Xu,
        Wenjing Jia,
        Tingcheng Cui,
        Ruomei Wang,
        Yuan-fang Zhang,
        and Xiangjian He
        % <-this % stops a space
\thanks{C. Xu, W. Jia, Y. Zhang and X. He are with Faculty of Engineering and IT, University of Technology Sydney, Australia (Email: \{Chengpei.Xu, Yuanfang.Zhang\}@student.uts.edu.au, Wenjing.Jia@uts.edu.au, xiangjian.he@gmail.com
). X. He will soon be with the Department of of Computer Science, University of Nottingham Ningbo China. X. He is the corresponding author.} 
\thanks{T. Cui is with Orbiseed Technology Inc., Toronto, Canada (Email: tingcheng@orbiseed.com).}
\thanks{R. Wang is with School of Data and Computer Science, National Engineering Research Centre of Digital Life, Sun Yat-sen University, China (Email: isswrm@mail.sysu.edu.cn).}

}

\maketitle

\begin{abstract}
One trend in the latest bottom-up approaches for arbitrary-shape scene text detection is to determine the links between text segments using Graph Convolutional Networks (GCNs).  
However, the performance of these bottom-up methods is still inferior to that of state-of-the-art top-down methods even with the help of GCNs. 
We argue that a cause of this is that bottom-up methods fail to make proper use of visual-relational features, which results in accumulated false detection, as well as the error-prone route-finding used for grouping text segments. 
In this paper, we improve classic bottom-up text detection frameworks by fusing the visual-relational features of text with two effective false positive/negative suppression (FPNS) mechanisms 
and developing a new shape-approximation strategy. 
First, dense overlapping text segments depicting the ``characterness'' and ``streamline'' properties of text are 
constructed and used in weakly supervised node classification to 
filter the falsely detected text segments. 
Then, relational features and visual features of text segments are fused with a novel Location-Aware Transfer (LAT) module and Fuse Decoding (FD) module to jointly rectify the detected text segments. 
Finally, a novel multiple-text-map-aware contour-approximation strategy is developed based on the rectified text segments, instead of the error-prone 
route-finding process, to generate the final contour of the detected text.  
Experiments conducted on five benchmark datasets demonstrate that our method outperforms the state-of-the-art performance when embedded in a classic text detection framework, which revitalizes the strengths of bottom-up methods. 
\end{abstract}


\IEEEpeerreviewmaketitle

\section{Introduction}

\maketitle

\IEEEPARstart{A}RBITRARY-SHAPE \add{scene text detection, localizing text instances with any shape, size and aspect ratio, is an upstream task of many downstream tasks such as text recognition~\cite{crnn}, document visual question answering~\cite{Ocr-vqa}, image retrieval~\cite{radenovic2018fine}, autonomous driving~\cite{Autonomousdriving}, layout analyse~\cite{Documentlayoutanalysis, xu2022semantic} etc.} Deep learning-based arbitrary-shape scene text detection methods generally follow one of two different pathways, 
\textit{i.e.}, top-down approaches and bottom-up approaches. 
Some top-down approaches 
consider a text instance as a whole and estimate its text area by leveraging the segmentation results, whereas the typical bottom-up methods tend to divide a text instance into text segments and then group them based on certain criteria. 
Both types of approaches~\cite{6,7,8,9,10,11,12} use a feature extraction network~\cite{vgg,resnet} with a Feature Pyramid Network (FPN)~\cite{5} as the basic text feature extraction network, and are deemed as the classic text detection framework.
The major difference is that 
these top-down approaches try to generate more accurate segmentation masks for text areas. 
Consequently, more sophisticated models for object detection and semantic segmentation have been adopted to enhance the performance of text segmentation. 
However, these methods may not be robust for long text instances due to their limited receptive field and the insufficient geometric feature representation of CNNs~\cite{16,7}.  

\begin{figure}[t]
  \centering
  \includegraphics[width=\linewidth]{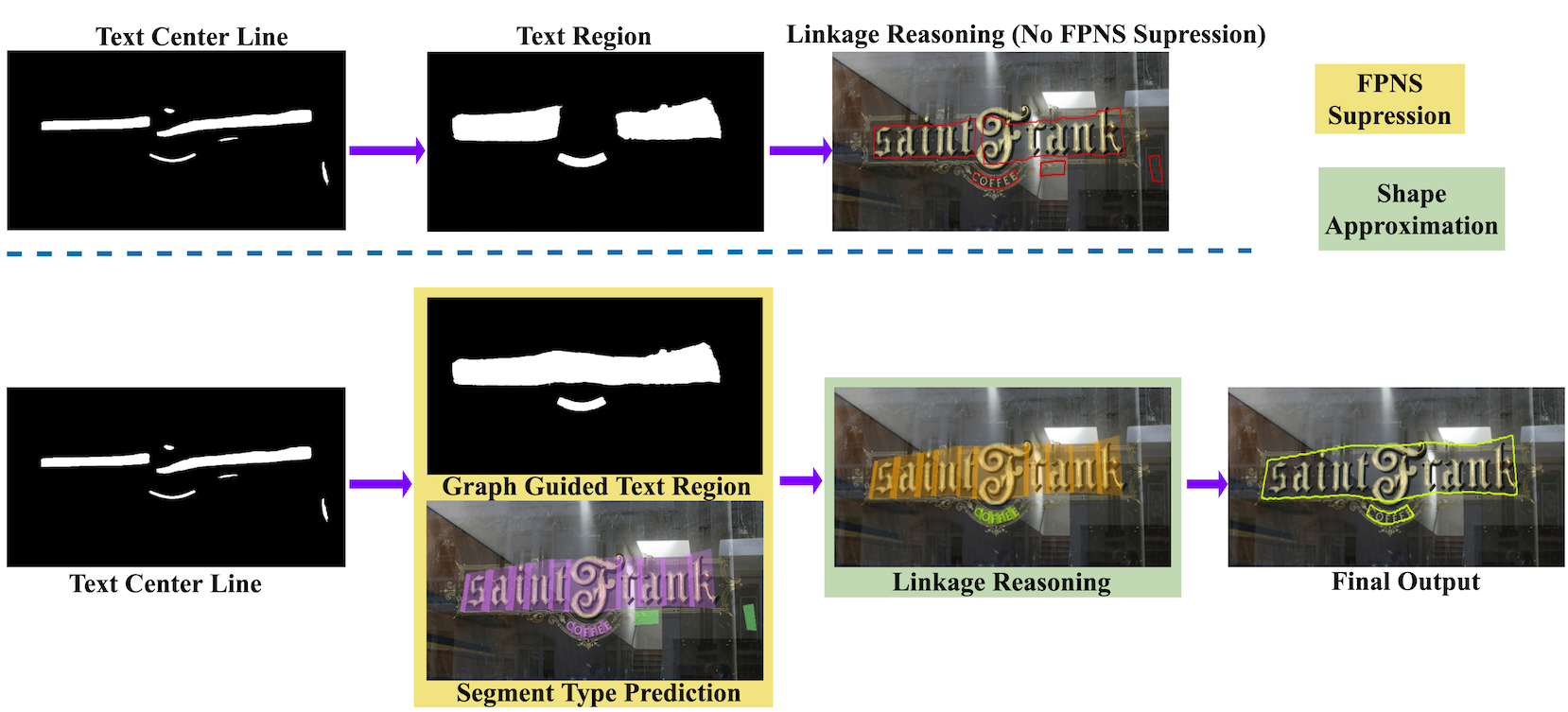}
    \caption{ \add{The error accumulation problem of existing bottom-up approaches (top) and our solution (bottom). Our Graph Guided Text Region fuses relational features with visual features and jointly determines the text area, which reduces false detections. 
    The segment type prediction module further rectifies false detections through excavating the ``characterness'' and connectivity of text segments.
    The final output shows that false detections such as missing parts and text-like objects have been suppressed.}} 
  \label{1}
\end{figure}

The bottom-up methods~\cite{7,8,11,6}
typically decompose a text instance into several components (\textit{i.e.}, text segments) and then group the eligible ones together with certain purpose-designed rules.  
Intuitively, these methods align better with how humans read text. 
Additionally, the bottom-up methods often require smaller receptive fields and are therefore more flexible for modelling arbitrary-shape texts. 
Therefore, they may be expected to yield higher accuracy and better robustness for long and curved texts. 
However, the reality is often the opposite, since bottom-up methods are also prone to accumulating intermediate errors (false positives/negatives). 
For example, in the challenging curved text datasets CTW1500~\cite{19} and Total-Text~\cite{18}, the overall accuracy of the best-performing bottom-up methods is lower than those of high-performing top-down methods, such as~\cite{9,43,10,42}. \add{This is because typical bottom-up methods usually have difficulty accurately determining the linkage, preventing error accumulation, and grouping text segments to obtain the final contour.}

\add{To understand the limitations of bottom-up methods, we need to understand the evolution of bottom-up methods in the deep learning era. }During this period, the relationship between text segments has become more diverse and complicated due to the focus of text detection research extending from horizontal text to quadrilateral text, and then to more challenging curved text. Existing bottom-up methods (\textit{e.g.},~\cite{24,6,21,25,3,12,xu2019lecture2note,xu2022morphtext},) with CNN, RNN and some pre-defined heuristic rules have considered visual similarity features, sequential features and geometric features in order to connect text segments.

However, the relationship between the text segments of arbitrary-shape text is non-Euclidean, which means there may still be connections between non-adjacent text segments especially in the case of curved texts. 
The Graph Convolutional Network (GCN) 
is an effective framework for representing this kind of non-Euclidean data, as it learns the representation of each node by aggregating the features from adjacent nodes~\cite{27}.  
In terms of capturing the relationship between text segments, state-of-the-art bottom-up methods such as DRRG~\cite{11}, ReLaText~\cite{7} and PuzzleNet~\cite{8} have adopted GCNs~\cite{27}.  
These methods share a similar framework, namely, a text segment proposal network  (\textit{e.g.}, VGG16/ResNet50+FPN) followed by a relational reasoning (link prediction) network (\textit{e.g.}, GCN). 
Nevertheless, the performance of these bottom-up methods is still lower than that of some high-performing top-down methods~\cite{9,43,tmm2}. 

\add{Besides the linkage relationship, there are two other problems that need to be addressed. }
First, the simple connection between the text proposal network and the GCN tends to result in error accumulation. 
An example is shown in Fig.~\ref{1}, where the wrongly separated text (false negatives) and two text-like objects (false positives) from the FPN layer of the text segment proposal network have been further propagated after the GCN based linkage prediction. Here, the Text Region/Text Center Line results generated from the same FPN layer share similar visual features and this property is often used as a double guarantee for accurate text area and text segment generation for further relational reasoning. 
However, sharing similar visual features sometimes leads to error accumulation. In this example, a text instance appears to be separated in both the text region map and its center line map as the FPN layer fails to build long-range dependency between the text segments in the middle. Although the relational information predicted by GCN can be treated as long-range dependency for different text segments, the simple connection of FPN and GCN does not take advantage of the relational features. Even if the text region and its center line are both considered for generating text segments, some of the text segment candidates can still be wrongly discarded, which significantly affects the accuracy of the GCN-based linkage prediction. 
The errors accumulated from text regions and text center line maps explain why the final detection result (the top-right figure) contains text instances that are incorrectly broken apart.
Hence, the reasoning of text regions needs stronger relational cues to ensure their connectivity and integrity.

Moreover, 
the simple connections between FPN and GCN and the weight sharing design in the FPN layer 
result in GCNs indiscriminately treating those false positive text segments the same as other text segments in the subsequent linkage reasoning steps, leading to falsely detected text-like objects. 
Among the GCN based bottom-up approaches, only ReLaText~\cite{7} has attempted to randomly generate some non-text nodes so as to remove the link with the non-text nodes. But this step is insufficient to suppress false positives/negatives, nor does it make advantageous use of the relational information.

Since the spaces between texts or characters are also non-text areas, both GCN and the text segment proposal network may be unable to distinguish text gaps from the non-text background.
As shown from the separated text instance in Fig.~\ref{1} (better viewed in darker colors), the separated area is right in the middle of the interval area of the text.  
As pointed out by~\cite{28}, lacking content information increases the possibility of false detection, indicating that too large~\cite{8} or too sparse text segments~\cite{7,11} cannot reflect the ``characterness'' of the content information of texts. 

Another problem occurs in the text segments' final grouping stage,  
\textit{i.e.}, the inference stage~\cite{6,11,21,24}, which also has a major impact on the final results but has not received much attention.
Some bottom-up methods (\textit{e.g.},~\cite{11,21,29,8,7}) need to locate or sample the points on the contour and determine the correct order in which these points are visited. 
This is because using the contour-drawing function such as \textit{cv2.drawContours}, the visiting order of contour points is critical for depicting a close contour and avoiding the edge crossing problem (a route-finding failure). %
It should be noted that the directional search towards the ends of text instances in TextSnake~\cite{21}, the sorting of the text segments' bounding boxes in TextDragon~\cite{29}, the searching of the polynomial curve that fits two long sides of text instances in ReLaText~\cite{7}, and the segmentation of the bisection lines based on the vertex extraction in PuzzleNet~\cite{8} can all be considered as 
different forms of searching for the visiting order of the contour points. 
This is analogous to the Traveling Salesman problem, which is an NP-complete problem. TextSnake adopted a complex heuristic rule-based route-finding method to alleviate this NP-complete problem. However,  
this post-processing may still end up with route-finding failure and hence is only applicable to specific conditions. 
The DRRG~\cite{11} used a greedy Min-Path algorithm to give an approximate solution to this problem. 
However, approximating solutions can sometimes fall into local optimal situations (see Fig.~\ref{2aa1}) when there are too many contour points. PuzzleNet and ReLaText adopted larger text segments to control the number of contour points so as to reduce the complexity of this NP-complete problem. 
However, larger text segments sometimes weaken the character level information of text and also compromise the flexibility and accuracy of the resultant text contours. 
How to find text contours without diminishing the ``characterness" of texts and introducing suboptimal failures needs to be further explored.

\begin{figure}[t]  
\subfigure[Route-finding failure from~\cite{11}]{
\begin{minipage}[t]{0.46\linewidth}

\includegraphics[width=3.8cm,height=2.5cm]{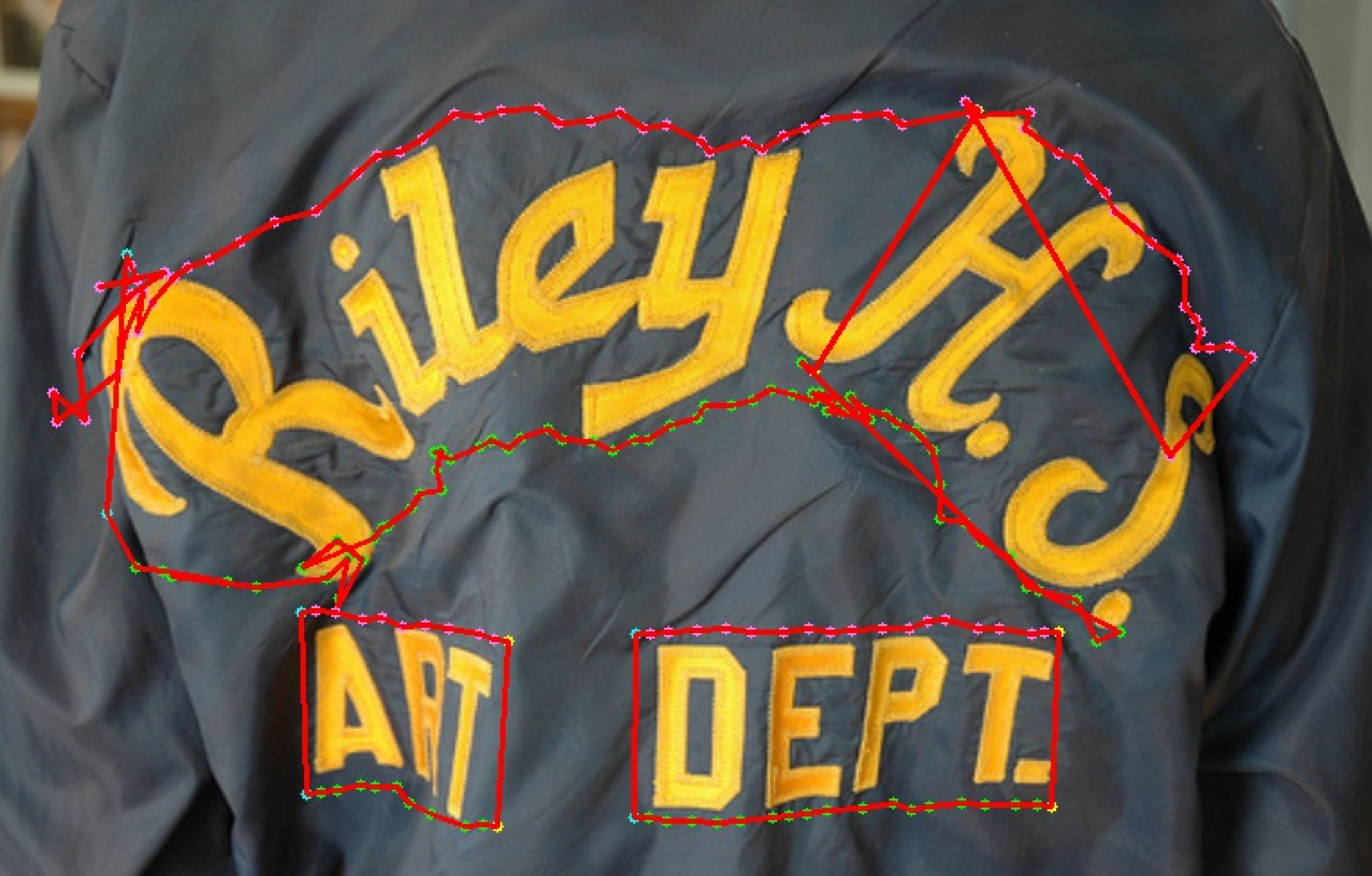}
\label{2aa1}
\end{minipage}
}               
\subfigure[Our shape-approximation solution ]{  
\begin{minipage}[t]{0.46\linewidth}

\includegraphics[width=3.8cm,height=2.5cm]{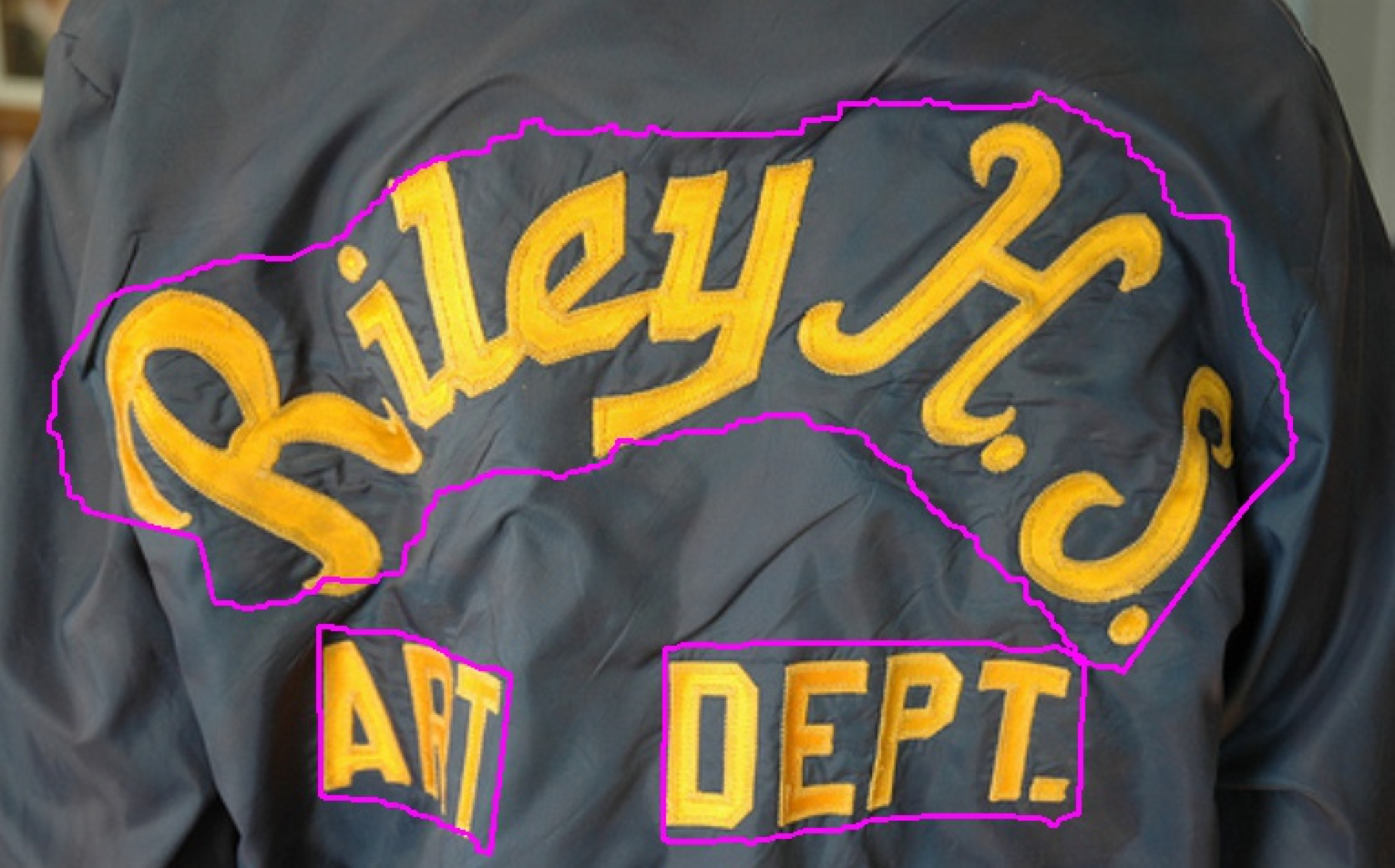}
\label{2b}
\end{minipage}   
} 
\caption{\add{Route finding may produce suboptimal visiting orders when there are too many contour points. The resultant crossing lines are caused by incorrect visiting order of intermediate points. These crossing lines may lead to self-intersecting contours and result in detection failures. }}
\label{2}
\centering
\end{figure}

In this paper, we propose designing the text segments 
in a dense and partially overlapping manner so as to retain the ``streamline'' characteristic of text, which allows flexible connection as much as possible to adapt text instances of arbitrary shapes. 
To ensure that the text segments are able to depict the ``characterness'' of text, text segments are designed to be fine enough to reflect characters and the spacings between them (see the segment type definition in Fig.~\ref{annotation}). 
Such a dense, overlapping design of text segments also 
benefits the construction of the subsequent 
fine-grained character-to-character graph structure,  
which enables character-level relational representation for further relational reasoning. 

Moreover, to address the simple connection and error accumulation issues, we 
re-examine GCN's other capabilities
to further classify the types of text segments. 
The existing methods~\cite{6,7,11} have used the linkage reasoning ability of GCN but ignored its node classification ability. 
When text segments are represented as nodes of a graph, segments of the same text instance are those who share similar “characterness” and “streamline” properties. 
The feature aggregation step in 
graph reasoning can also differentiate true positive text segments by relational reasoning. This establishes a backward feedback mechanism that can retrieve and suppress false detections (including both false positives and false negatives) during link prediction. Classifying the types of segments enables the GCN to examine the content information 
for more accurate text region prediction. 
This forms one of our two false positive/negative suppression (FPNS) strategies. 

Secondly, we develop another FPNS strategy that aggregates the text's visual and relational features to suppress false detection, as shown in Fig.~\ref{1}. 
Instead of considering only visual text feature maps from the output of FPN layers, here we propose 
a Location-Aware Transfer (LAT) module to convert relational features produced by GCNs into visual compatible features. Then, a Fusion Decoding (FD) module is introduced 
to fuse the relational features with the visual features to generate a Graph Guided Text Region (GGTR) map. Since relational features are a ready-made high-dimensional representation of long-range dependency based on the relational connection of text segments, the GGTR map provides additional long-range dependency to guide the connectivity and integrity of proposed text regions and suppress false detection, resulting in more potential true positive text segments. 
The GGTR map forms the other FPNS strategy to ensure enough candidate text segments are generated for further graph reasoning.
Thus, the two FPNS strategies decompose the overall text map rectification problem into the problem of rectifying text segments to improve their fault tolerance.

Finally, 
instead of the error-prune route-finding process, we 
develop a shape approximation (SAp) strategy to
group the rectified text segments and approximate the contour of the text.
Such a design allows an unlimited number of text segments to approximate the contours of arbitrary-shape text.

In summary, the main contributions of our work are as follows.

1) We utilize the relational feature of GCNs to rectify text segments by globally considering their ``characterness'' and ``streamline'' in the same relational structure through a weakly supervised training process. 
To the best of our knowledge this is the first time the classification ability, instead of link prediction, of GCNs is used for scene text detection. 

2) We propose a novel visual-relational reasoning approach to \add{increase the feature discriminability for falsely detected text segments} in typical bottom-up arbitrary-shape text-detection approaches and take advantage of their strengths. This is demonstrated to be effective in capturing both visual 
and continuity properties of the characters that determine text areas.

3) 
We redesign the text segments 
in a dense and partially overlapping manner 
and develop a simple but effective contour inference strategy to depict the ``characterness'' of text, which can handle complex situations of arbitrary-shape text with enhanced relational reasoning and type classification capabilities.

Experiments conducted on several curved text detection benchmark datasets show that 
our approach surpasses the state-of-the-art performance and demonstrates the strength of bottom-up approaches for arbitrary-shape scene text detection. 
This shows that bottom-up methods are not necessarily inferior to, but can be even better than, top-down methods in arbitrary-shaped text detection.

\section{Related Work}

\subsection{Bottom-up Text Detection Methods} 
Typical bottom-up methods divide text into segments and then reason the link between them.
Following this design structure, CTPN~\cite{24} was a pioneering work that brought text detection into the deep-learning area, but it simply merges text segments according to a certain threshold and could only deal with horizontal texts. SegLink~\cite{3} was designed to detect multi-oriented texts, aiming to connect the centers of two segments with an eight-neighbourhood link prediction. 
Subsequently, as the research focus changed to curved text, text detection transitioned from the detection of bounding boxes to the detection of contours. 
TextSnake~\cite{21} was the first work to attempt this transition. It used several circles with predictable angles and radius to fit curved texts. 
At this point, both geometric and visual information have largely been considered in curved text detection.

To further establish the relationship between text segments, Baek \textit{et al.}~\cite{6} used affinity boxes to describe the neighbouring relation between text segments for further spatial reasoning, but it was insufficient for reasoning the complex relationship between curved texts and reflecting the ``characterness'' of texts. 
Later, researchers used GCNs to integrate geometric features, visual features, and relational features for link prediction~\cite{11,7,8}.  
These three methods shared a similar framework but differed in the size and number of text segments. 
PuzzleNet~\cite{8} used the largest text segments to merge adjacent text segments with angle differences less than $\pi/36$, which was inflexible in the following segment grouping stage. 
RelaText~\cite{7} generated smaller text segments than PuzzleNet, but they were still too large and too sparse to depict the ``characterness'' feature of the text. 
The size of the text segments in DRRG~\cite{11} was the smallest, and although it contained more character-level features and was more flexible when grouping, its route-finding process was of significantly higher complexity. Thus, there is often a trade-off between the size and number of text segments and the difficulty of the grouping operation, which needs to be further explored. 

In our approach, we use ``characterness'' and ``streamline'' to depict the character-level and adjacency-level features of text content and a shape approximation (SAp) strategy to avoid the computational overload of the route-finding process while greatly increasing flexibility.

\subsection{False Detection Suppression}

Suppressing false detection requires distinguishing text and non-text areas accurately. 
A first prior approach is to strengthen the network's capability for depicting text features by adding additional text content information.  
Some methods try to use character-level annotation to increase the ``characterness'' expression. For example, CRAFT~\cite{6} adopted a weakly-supervised training strategy to train the character region score to split words into characters;
TextFuseNet~\cite{9} fused word-level, character-level and global-level features of text to conduct text instance segmentation using Mask-RCNN~\cite{13}.  
Similarly, Zhang \textit{et al.}~\cite{20} also used this framework to obtain character-level annotation with a weakly-supervised training strategy. However, training character-level annotation as an instance segmentation task means doing detection and recognition together, which brings additional computational burden. 
Other methods such as~\cite{8,20,36,7,39,43,tmm3,tmm1} attempted to insert blocks like Non-local~\cite{35}  or Deformable Convolution~\cite{37} to increase the network’s capability for extracting text features, but detection results showed that these methods often resulted in limited improvement. 
We need to focus more on the characteristics of the text itself. 

A second prior approach is to jointly consider multiple text maps to get a final text area. ContourNet~\cite{10} tried to suppress non-text areas by considering text maps in horizontal and vertical directions, which effectively suppressed false positives. However, their method still suffered from false negatives when both horizontal and vertical text maps had defects. 
TextSnake~\cite{21} and DRRG~\cite{11} predicted an additional Text Center Line (TCL) map and multiplied it with the original Text Region (TR) map for FPNS}.
As shown in Fig.~\ref{1}, the FPN produces both TCL and TR maps. It is very likely that both TCL and TR maps are flawed, which may result in error accumulation.
Moreover, simply setting thresholds on different text maps~\cite{10} or directly multiplying different text maps 
increases the risk of suppressing the true positive text areas as well. What is required is a more effective and accurate approach to consider the feature of text instances and indirectly use text maps.

\section{Methodology}
\begin{figure*}[t]
 
  \includegraphics[width=\linewidth]{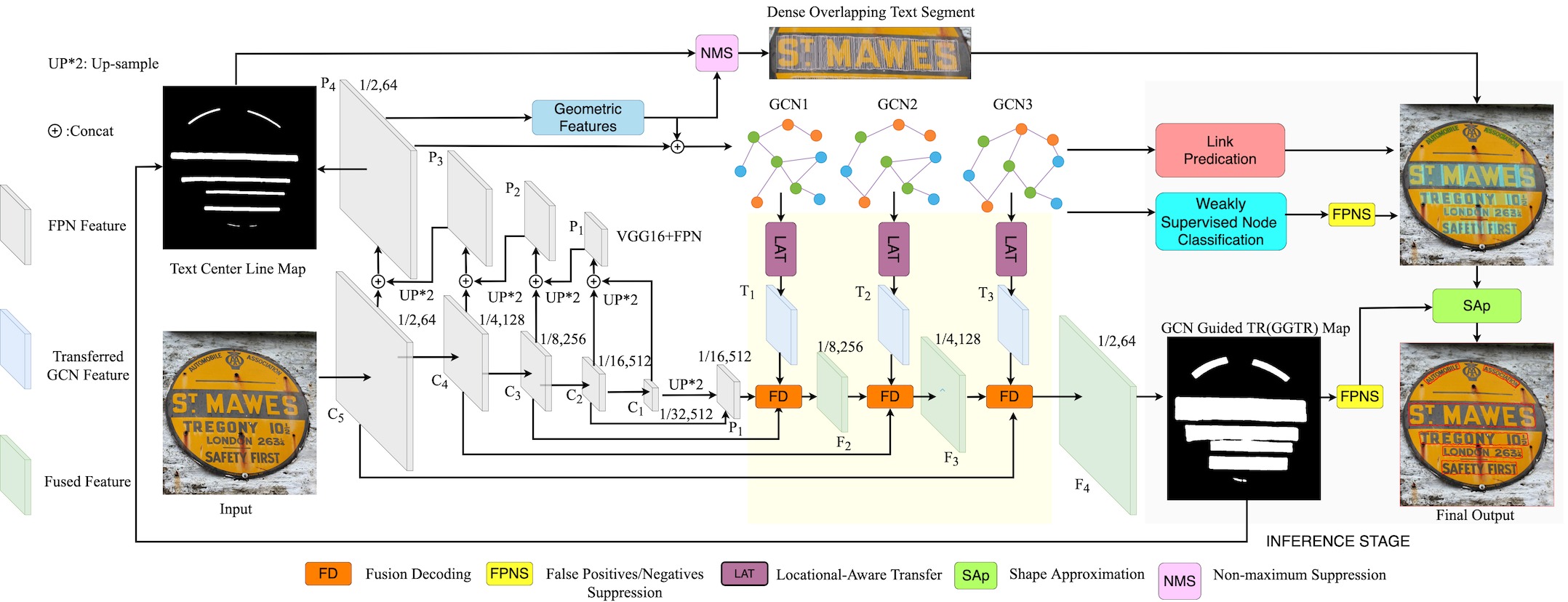}
  \centering
  \caption{\add{The overall structure of our network. The ``1/2,64", ``1/4,128",... and ``1/16,512" indicate the scale ratio and the channel number of the input image. In the training flow, the TCL map, GGTR map, geometric features, link prediction and node classification are guided by the loss function. In testing flow, the GGTR map and weakly supervised node classification results are used in our proposed FPNS strategies to rectify text segments. The proposed SAp is used to obtain the final contour of text segments.}}
  \label{3}
\end{figure*}

The overall structure of our network is illustrated in Fig.~\ref{3}. We adopt VGG16 plus FPN for visual feature extraction. 
Visual features $C_5$ to $C_1$ are first extracted and up-sampled to $P_1$ to $P_4$. 
$P_4$ is used to generate TCL maps and geometric features of possible dense overlapping text segments. $P_4$ and the geometric features are embedded into nodes and then converted into a graph structure. 
Then, the text segment types are annotated by a weakly supervised network. 
The details of the text segment generation, annotation steps and weakly supervised training are provided in Sect.~\ref{Text Segment Generation}. 
Sect.~\ref{Graph based Reasoning and Fusion} describes the deployment of 
three GCN layers to learn %for learning 
the graph representation between text segments, where the linkage relationship and the text segment types are reasoned by aggregating the features of adjacent nodes in the graph structure.

Meanwhile, in Sect.~\ref{fd+lat} the visual feature $P_1$ is enhanced by 
aggregating with 
the relational feature produced by GCNs to provide additional long-range dependency. 
Since visual feature and the relational features have different dimensions, we develop firstly a Location-Aware Feature Transfer module to align the two types of features and secondly a Fusion Decoding (FD) module to fuse them. 
Finally, we obtain a Graph Guided Text Region (GGTR) map based on the visual-relational features.  

During the training stage, the training of the TCL map, GGTR map, geometric features, link prediction and node classification are guided by the loss function designed in Sect.~\ref{loss}. 
During the inference stage (see Sect.~\ref{inference} for details), the GGTR map that contains visual-relational features is used to rectify the TCL map and contour generating process for suppressing false positive/negative text segments. 
The weakly supervised node classification results are used to further rectify the false positive/negative text segments. 
Finally, a novel text-instance-based contour inference module is used to approximate the contour of the rectified text segments to obtain the final results.

\subsection{Dense Overlapping Text Segments}

Text segments are basic components of many 
bottom-up methods 
and have therefore been used in existing works. 
In this paper, our re-designed dense overlapping text segments play an important role in our method, since they are involved in the generation of graph structure that is used in GCNs and the proposed shape approximation strategy. 
In the following sub-sections, we first give the details of the text segments, 
then describe how the ground-truth type of each text segment is obtained 
in a weakly supervised manner. 

\subsubsection{Text Segment Generation}
\label{Text Segment Generation}

We first adopt VGG16+FPN for generating the TCL map, as well as the geometric features for each text segment. The text segments are defined as small rectangles by $(x, y, h, w, \theta)$, where $x$ and $y$ represent the $X$ and $Y$ coordinates of the rectangle's center point, and $h$, $w$ and $\theta$ represent the height, width and rotation angle of the rectangle, respectively. 

As shown in Fig.~\ref{3}, 
the final extracted features $P_4$ are used to generate the TCL map, and the geometric features (\textit{i.e.}, Height map $H$
and Angle map $\Theta$) are used to restore the geometric representation of text segments.  
More specifically, $Conv_{1\times1}$ is applied on $P_4$ to obtain these text maps, namely, TCL, H and $\Theta$. 
The Width map (W) is obtained by applying a clip function to the Height map while keeping the width $w$ between 2-6 pixels, as shown in Fig.~\ref{4}.

Note that, the width $w$ of text segments affects  the generation of the graph structure and the final contour shape, and hence influences the overall performance of text detection. 
This is because our constructed visual representation for each text segment is designed to depict the ``characterness'' and ``streamline'' of the text instance. 
Since characters in a text instance are usually taller than they are wide, denser text segments with a smaller width can therefore approximate the contour of text instances more accurately using their connected shapes. 
Also, smaller text segments better differentiate characters, character spacing and word spacing in the text, which will be used to further suppress false detections. 
The ablation studies conducted in Sect.~\ref{ablation} demonstrate this with extended discussion on the impact of the width $w$ on the final results.

Moreover, the ground truth of the TCL map is obtained 
by shrinking the ground truth text area vertically 
from top and bottom towards the middle line 
using the method described in~\cite{11,21}. 
To generate dense overlapping text segments, here we first generate small rectangles 
centered at each text pixel of the TCL map,  
with their height, width and angle based on the Height, Width and Angle maps
\footnote{It should be noted that in the inference stage described in Sect.~\ref{inference}, the TCL map will be rectified by the GGTR map before generating dense overlapping text segments.}
Then, a non-maximum suppression (NMS) algorithm with an IoU threshold of 0.5 is applied on this TCL map to remove those heavily overlapping rectangles. 
This step balances the computational load and ensures the density and connectivity of the text segments, which increases the flexibility for them to fit the contours of text of various shapes. 
 
\subsubsection{Weakly Supervised Text Segment Annotation}
\label{annotation}

\begin{figure}[t]
  \setlength{\abovecaptionskip}{0.2cm}
  \centering
  \includegraphics[width=8cm]{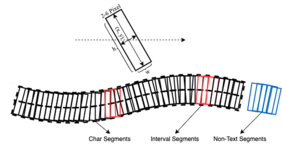}
  \caption{\add{The three types of text segments. All of these text segments are designed to have overlap with each other with a width between 2-6 pixels.}}
  \label{4}
\end{figure}

After we have obtained the dense overlapping text segments, the next step is to determine their types. 
The type information of text segments enables them to better reflect text's “characterness” and allows building a character-to-character graph structure for further fine-grained relational reasoning and rectification.
In our work, text segments are categorized into one of three types: Char Segments, Interval Segments and Non-Text Segments (see Fig.~\ref{4}). 
Char Segments contain characters or partial characters only, Non-Text Segments are background, and Interval Segments are the gaps or spacing between characters or words.

Interval Segments can often be confused with Non-Text Segments, especially in text instances with large inter-character spaces, so they need to be classified under close supervision. 
However, none of the real-world datasets provide annotation for the intervals between characters. 
Therefore, in our approach we utilize a synthetic dataset ``SynthText''~\cite{synthetic} that has both character-level and word-level annotations, 
and propose a weakly-supervised training approach to classify the type of text segments. 

To start the weakly supervised learning, labelled samples are first generated from the SynthText dataset. 
Specifically, 
denote the $i$-th text segment with the center point coordinates 
$(x_i,y_i)$ as $S^i$, 
the collection of all text segments as $S^*$, 
the collection of Char Segments as $S_{char}^*$, the non-character segment set $S^*-S_{char}^*$ as $S_{\neg char}^*$, 
the collection of all Interval Segments as $S_{inter}^*$,  
and the union of all character areas in the image as $C$. 
With these initial samples, we can train a preliminary node classifier, 
which is then fine-tuned on real text datasets with only word-level annotation to learn the types of text segments in an iterative, weakly-supervised manner.

During the iterations of the weakly-supervised learning, 
the node classifier may misclassify some Char Segments as Interval Segments. 
Subsequently updating the loss based on these false Interval Segments will confuse the training and seriously affect the performance of the learned model. 
Therefore, we consider text segments' adjacent segments and use a simple approach to ensure 
only those more reliable Interval Segments back-propagate the loss.

Towards this, we first define the distance between each Char Segment in $S_{char}^*$ and non-char segment in $S_{\neg char}^*$ as the Euclidean distance of their centers:
\begin{equation}
d(S_{char}^i,S_{\neg char}^j)=\left\Vert(x_i,y_i)-(x_j,y_j)\right\Vert_2, 
\end{equation}
where $\forall S_{char}^i \in S_{char}^*$ and $\forall S_{\neg char}^j \in S_{\neg char}^*$.

\begin{figure}[t]  
\subfigure{
\begin{minipage}[t]{0.46\linewidth}

\includegraphics[width=4.2cm,height=2.5cm]{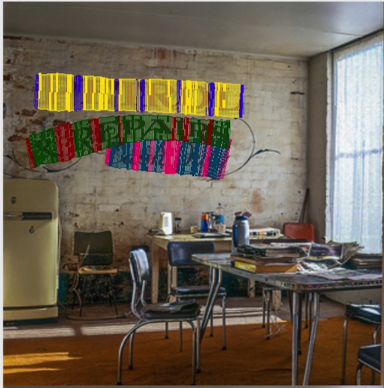}
\label{2aa}
\end{minipage}
}               
\subfigure{  
\begin{minipage}[t]{0.46\linewidth}

\includegraphics[width=4.2cm,height=2.5cm]{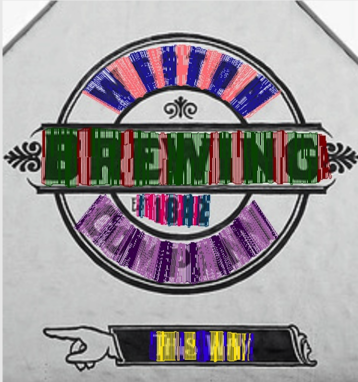}
\label{2b}
\end{minipage}   
} 
\vskip -16pt
\subfigure{
\begin{minipage}[t]{0.46\linewidth}

\includegraphics[width=4.2cm,height=2.5cm]{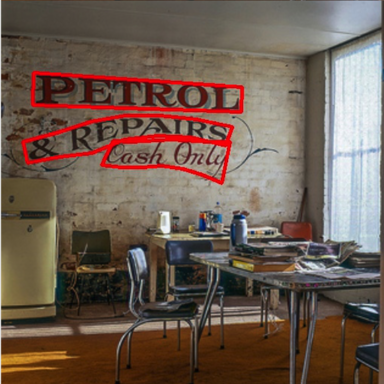}
\label{2aa}
\end{minipage}
}               
\subfigure{  
\begin{minipage}[t]{0.46\linewidth}

\includegraphics[width=4.2cm,height=2.5cm]{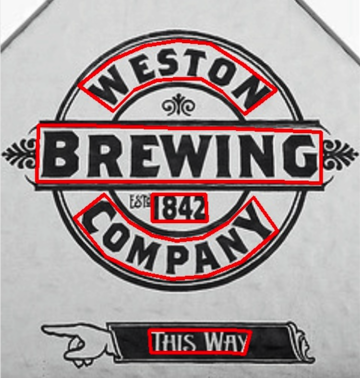}
\label{2b}
\end{minipage}   
} 
\vskip -16pt
\caption{The results of weakly supervised annotation (1st row) and the ground truth (2nd row). The Interval Segments and Char Segments are shown with different colors. The GCN now has the ability to classify the type of segments, which will benefit the following FPNS step. In these figures, Non-Text Segments are not displayed for clarity.}
\label{char}
\centering
\end{figure}

To alleviate the situation of updating the loss based on these false Interval Segments, 
for each Char Segment $S_{char}^i$, we only perform gradient update for its closest Interval Segment $S_{inter}^i$, which is the non-char segment that has the minimal Euclidean distance from $S_{char}^i$, \textit{i.e.}, 
\begin{equation}
\!S_{inter}^i\!\!=\!\left\{{S_{\neg char}^k \big| d(S_{char}^i,S_{\neg char}^k)\!=\!\!\!\!\!\!\! \min_{\!j\leq|S_{\neg char}^*|\!}\!\!\! \!d(S_{char}^i,S_{\neg char}^j)\!}\right\}.
\label{interval}
\end{equation}
It is worth noting that, here although only the closest non-char segments are defined as the Interval Segments go through the gradient update process, those non-char segments further away from the Char Segments may still have a high chance of being correctly classified as Interval Segments after the weakly supervised training.
The Interval Segments shown in Figs.~\ref{1} and~\ref{char} demonstrate this.
 
Furthermore, 
since there is no char-level ground-truth or internal ground-truth available in the existing real text datasets, in our approach we use a simple mechanism to determine whether a weakly supervised training result is acceptable: the union of the predicted Char Segments and Interval Segments covers at least 90\% of the annotated text areas and the annotated text areas do not contain Non-Text Segments. 
Thus, with the iteration of training, gradually more and more text segments are correctly classified until most classified Char Segments and Interval Segments are in annotated text areas.

With the weak supervision it is likely that some Interval Segments are misclassified as Char Segments (and vice versa), especially when the characterness properties of the characters are rather weak (\textit{e.g.}, blurry, tiny, etc). 
However, our main concern is the correctness of distinguishing Non-Text Segments and Interval Segments to prevent the text instance being mistakenly broken apart due to the misclassification of some Interval Segments. 
The exemplar image in Fig.~\ref{3} (see the area between ``Y'' and ``10'' in the image) shows an example of a large interval space, which, if misclassified may result in the text instance being broken apart.
Therefore, as long as they are not misclassified as Non-Text Segments (\textit{i.e.}, background), such misclassification will not cause errors, as our FPNS strategies only suppress those segments classified as Non-Text Segments.

\subsection{Graph based Reasoning}
\label{Graph based Reasoning and Fusion}

We now give details about 
building the relational representation between text segments, as well as link prediction and node classification, after the dense overlapping text segments have been obtained.

To predict the links between text segments and classify their types, %the type of each text segment, 
we first need to integrate the feature representation of each text segment and construct their graph structures. 

We consider two types of features for text segments, \textit{i.e.}, the visual feature $F_{vis}$ 
\add{is obtained by applying convolutional operation on $P_4$ in FPN for generating the TCL map and the geometric feature $F_{geo}$. }
We note $F_{seg}$ as the feature representation matrix of text segments. 
Following the methods in~\cite{8,11}, the RRoI Align~\cite{38} is adopted to pool the visual features from FPN. 
$F_{geo}$ is formed by concatenating the feature maps $TCL$, $H$, $W$ and $\Theta$, and is embedded into the same space as $F_{vis}$ as: 
\begin{equation}
F_{seg}=[F_{vis}(P_4); F_{geo}([TCL;H;W;\Theta])],
\end{equation}
where the $[;]$ operator indicates concatenation.

To build the graph structure of text segments, we treat each text segment as the pivot and connect its eight closest text segments (measured by the Euclidean distance of their center points) as its 1-hop nodes. 
Then, for each 1-hop node, we connect its four closest text segments (excluding the pivot). 

The pivot, its 1-hop nodes and 2-hop nodes compose the basic graph structure for the text segment, which can be represented by an adjacency matrix $\mathcal{A}$. We consider the top three closest text segments of a pivot to have a link with the pivot. Similar to those in~\cite{41,27,11},
the graph convolution is represented by:
\begin{equation}
F_{seg}^{(i+1)}=\sigma([F_{seg}^{(i)}; \tilde{\mathcal{D}}^{\frac{1}{2}} \tilde{\mathcal{A}} \tilde{\mathcal{D}}^{\frac{1}{2}} F_{seg}^{(i)} ] W^{(i)}),
\label{con:7}
\end{equation}
where the $F_{seg}^{(i+1)}$ is the feature representation for the $i+1$ layer after the graph convolution of $F_{seg}^{(i)}$. $ \tilde{\mathcal{D}}^{\frac{1}{2}} \tilde{\mathcal{A}} \tilde{\mathcal{D}}^{\frac{1}{2}}$ is the re-normalization trick from the original GCN paper~\cite{27} and can be calculated by $\mathcal{A}$. $\sigma$ is a non-linear activation function and $W^{(i)}$ is the weight of the graph convolutional layer.

Thus, the relational link prediction can be written as:
\begin{equation}
\hat{r}=Softmax(m^{(i+1)}(F_{seg}^{(i+1)})),
\label{7}
\end{equation}
where $m^{(i+1)}$ denotes a multi-layer perceptron with a PReLU activation. 
The feature matrix $F_{seg}$ can also be written as $F_{seg}=[f_1,f_2,...f_v,...]^T$, where $f_v$ is the feature representation vector for the single text segment $v$. 

The GCN segment classification results for the text segment $v$ can be represented by:
\begin{equation}
\hat{f_v}=Softmax( \varphi (\frac{1}{|\mathcal{N}(v)|} \sum_{u \in \mathcal{N}(v)} \mathcal{W}f_{u}+b)               ),  \forall v \in \mathcal{V}, 
\label{con:8}
\end{equation}
where $\mathcal{N}(v)$ means the neighbor segments in the graph structure $\mathcal{V}$ of text segment $v$, and $\mathcal{W}$ and $b$ are the weights of the graph convolution. $\varphi$ is the non-linear activation function. Eq.~\eqref{con:8} is the node-level graph feature aggregation of Eq.~\eqref{con:7}.

After feature aggregation, 
the relational reasoning results can be used to combine text segments that have a strong relationship. The weakly supervised node classification results are used to refine the detection results inherited from the previous FPN layer. 
Thus, the text segment type reasoning prevents detection errors from being further accumulated. 
Both link prediction and node classification 
make inferences based on the high-dimensional relational features on the same local graph, and can benefit each other when sharing the same weights. 
The node classification steps further classify Text Segments and Non-Text Segments, reducing the number of false text segments being incorrectly linked %by the 
in link prediction.% step. }
\\

\subsection{Visual-Relational Feature Fusion}
\label{fd+lat}

The relational features from GCN layers provide a ready-made high-dimensional representation of long-range dependency 
depicting the relational connection of text segments.
We fuse them with the visual features to complement the long-range dependency between text segments for generating graph-guided text regions. 
However, the dimensionality  of the relational features is different from that of the visual features, which means they cannot be fused directly. 

To address this issue, we propose Graph features' Location-Aware Transfer (LAT) to reconstruct the graph convolutional feature of $i$-th GCN layer $F_{seg}^i$ in Eq.~\eqref{con:7} with consideration of the location information of each node in the graph. The relational features after GCN contain the structural information of the graph in its first two dimensions.
For example, the relational feature $F_{seg}^i \in \mathbb{R}^{N\times M \times D}$, where $N$ is the total number of nodes (\textit{i.e.}, text segments) in the input image, $M$ is the number of 1-hop nodes, and $D$ is the dimension of GCN features which is similar to the channel space in the convolutional layer. Our goal is to reshape the $F_{seg}^i$ of dimension ${N\times M \times D}$ to ${C\times H \times W}$, where $C$, $W$ and $H$ are the channel, width and height of the input image.
The core idea is to fill in the graph features in the corresponding region of the transferred map based on the location information of each text segment. Namely, building connections of the nodes on the graph structure with their location on the image, as the location of each node is also translation invariant compared to the visual feature.
The transferred feature $T_i \in \mathbb{R}^{C\times H \times W}$ for $F_{seg}^i$ after LAT can be calculated with Algorithm~\ref{ag0}. 
\begin{algorithm}
\footnotesize
\caption{\add{Pseudocode of} Location-Aware Transfer (LAT)}
\label{ag0} 
\KwIn{$F_{seg}^i$, $F_{geo}$ }
\KwOut{Transferred feature $T_i$}
 $T_i=ZeroLike(H,W,C)$\\
 $N=F_{seg}^i.shape[0]$\\
\For{ $ k \leq N$ }{
    $(x_k,y_k,h_k,w_k,\theta_k)=Geometric(F_{geo},S^k)$\\
    \If{$type(S^k)\neq nontext$}{
           \For{$\forall x,y \in bbox(x_k,y_k,h_k,w_k,\theta_k)$}{
                    $T_i[x,y,:]=F_{seg}^i[k,:,:].flatten()$\\
           }
    $T_i=T_i.transpose(2,0,1)$
    }
}
{\bfseries return $T_i$}
\end{algorithm}

First, we calculate the geometric feature $(x_k,y_k,h_k,w_k,\theta_k)$ of text segment $S^k$ from the geometric feature map $F_{geo}$. Here, we only consider transferring the graph features of Char/Interval Segments. Then, we can find the bounding box (bbox) of each text segment on the feature map of $T_i$ according to  $(x_k,y_k,h_k,w_k,\theta_k)$. For each point ($x,y$) in the bbox, we fill the flattened graph feature into $T_i$. 

\begin{figure}[t]
 
  \includegraphics[width=\linewidth]{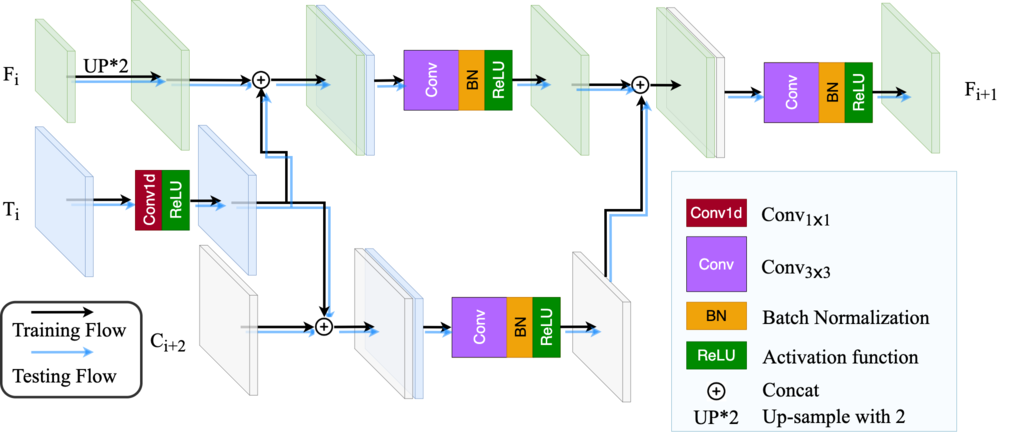}
  \centering
  \caption{\add{Network structure of the multi-modal fusion decoding (FD) module.}}
  \label{fig:model}
\end{figure}

Since FPN is unable to reason the relationship between different text segments, we introduce a multi-modal Fusion Decoding (FD) module to capture the long-range dependency between individual text regions according to their relational features as well as FPN features. The final goal is to generate a Graph Guided Text Region (GGTR) map for further FPNS. 

As shown in Fig.~\ref{fig:model}, the proposed FD module produces the final fused feature $F_{i+1}$ as:
\add{
\begin{equation}
F^{\prime}=\left\{
\begin{aligned}
CBR[UP(P_{i});CR(T_i)]\quad {\text {if}} \quad i=1 \\
CBR[UP(F_{i});CR(T_i)]\quad\quad\quad\quad\quad \\
\end{aligned}
\right.
\end{equation}
\begin{equation}
F_{i+1}=CBR[F^{\prime};CBR[(C_{i+2}); CR(T_i)]],
\label{con:9}
\end{equation}}
where $CBR$ denotes $Conv_{3\times3}$, Batch Normalization and ReLU, \add{$CR$} donates $Conv_{1\times1}$ and ReLU, and \add{$UP$} denotes two times of up-sampling. Finally, we apply the $Conv_{1\times1}$ on $F_{3}$ to obtain a GGTR map, which will be used as a visual-relational guide in the inference stage.

\subsection{Loss Function}
\label{loss}
The overall objective function consists of six parts and is formulated as:
\begin{equation}
L=L_{GGTR}+L_{TCL}+L_{H}+L_{\theta}+L_{Link}+L_{Node}, 
\end{equation}
where $L_{GGTR}$ and $L_{TCL}$ are the cross-entropy losses from the GGTR and TCL maps, respectively, and $L_{H}$ and $L_{\theta}$ are Smoothed $L1$ losses for Height and Angle.
We adopt OHEM~\cite{40} for training $L_{TCL}$, and the ratio of positive sample and negative sample areas is set to $1:3$. 
The linkage loss $L_{Link}$ is the cross-entropy between the ground truth linkage relationship of text segments and $\hat {r}$ in Eq.~\eqref{7}. 

For the text segment classification loss $L_{Node}$, since our annotation is obtained with weakly supervised training, we only consider the text segments that are labeled as Char Segments, Interval Segments and Non-Text Segments (denoted as $f^{v}_c$, $f^{v}_i$, and $f^{v}_n$ respectively) and ignore those that have not been labeled. Thus,
\begin{equation}
L_{Node}=-(\sum_{v \in \mathcal{N}}{f^{v}_c \log{\hat{f^{v}_c} }}+\sum_{v \in \mathcal{M}}{f^{v}_i \log{\hat{f^{v}_i} }}+\sum_{v \in \mathcal{O}}{f^{v}_n \log{\hat{f^{v}_n} }}),
\end{equation}
where $\mathcal{N}$, $\mathcal{M}$, and $\mathcal{O}$ are the collections of all Char Segments, Interval Segments and Non-Text Segments, respectively.

\subsection{Inference by FPNS and Shape-Approximation (SAp)}
\label{inference}
The Graph Guided Text Region map and the text segment classification results from the GCN are then used to rectify the TCL map to remove false detection. 
The relationship prediction and the dense overlapping text segments together ensure the completeness and accuracy of using the contour of the grouped text segments to approximate the contour of the text. The connection of the text areas is ensured by the ``characterness'' and ``streamline'' of our dense overlapping text segments.
The overlapping of text segments also reflects their connectivity, ensuring that the shape of the grouped segments can be used to approximate the contour of the original text instances accurately. 
In our work, instead of jointly rectifying multiple text maps as in~\cite{10,11,21}, we decompose the FPNS problem based on the grouped text segments reasoned by visual-relational features. Therefore, our approach does not use element-wise multiplication or search the entire map, which avoids accidentally removing true positive text areas. 

\begin{algorithm}

\caption{\add{Pseudocode of} Inference with FPNS and shape-approximation  }
\label{ag} 
\footnotesize
\KwIn{Image, GGTR map, TCL map, H map, W map, $\Theta$ map, $GCN_{node}$ and $GCN_{link}$}
\KwOut{Text Contours $\mathcal{C}$}
$\mathcal{C}=List()$\\

$TCL^{\prime}=TCL \bigcap Shrink(GGTR)$ \algorithmiccomment FPNS (GGTR)\\
$TCLContours=cv2.findContours(TCL^{\prime})$\\
$\mathcal{S}=List()$\\
\For{$contour  \in TCLContours $ }{
    $Segments =Getboxes(TCL^{\prime}, H, W, \Theta, contour) $\\
    $Segments=NMS(Segments, 0.5)$ \\
\For{$seg  \in Segments$}{
    $type=GCN_{node}(seg)$\\
    \If{$type == nontext$}{
         $Segments.remove(seg)$ \algorithmiccomment FPNS (Node) 
    }
    $\mathcal{S}.append(Segments)$
}
}
$\mathcal{\hat{S}}=GCN_{link}(\mathcal{S})$ 

$Blankimage=ZeroLike(Image)$\\
\For{ $ConnectedSegments  \in \mathcal{\hat{S}}$ }{
     \For{$box  \in ConnectedSegments$}
     {    
        $cv2.drawContours(Blankimage,box)$
    }
$Blankimage=(Blankimage \oplus B_{3\times 3}) \ominus B_{3\times 3}$\\
$Contour=cv2.findContours(Blankimage)$\\
\If{$IoU(Blankimage,Blankimage*GGTR)>= 0.5$}{
        $\mathcal{C}.append(Contour)$\algorithmiccomment  FPNS (GGTR)\\
    }
}
{\bfseries return} $\mathcal{C}$
\end{algorithm}
Algorithm~\ref{ag} illustrates our proposed shape-approximation algorithm. 
The inputs require the text maps of GGTR, TCL, H, W and $\Theta$, as well as node classification $GCN_{node}$ and link prediction components $GCN_{link}$ from the GCN.
First, we perform a union operation on the TCL map and shrink the GGTR map to produce a rectified TCL map $TCL^{\prime}$. The text area of the GGTR map is shrunk from the top and bottom towards the middle until single-pixel height is reached. With this step, the GGTR map is introduced to provide the long-range dependency of text segments and also provide more candidates for the graph reasoning stage.
The function $cv2.findContours$ is used to find the potential text center lines of the text instances. Following that, for each text instance, the $Getboxes$ function is used to generate dense text segments for pixels in the rectified TCL map. 
The classifications from $GCN_{node}$ are used here to suppress false detection of text segments, and the remaining text segments are stored in the list $\mathcal{S}$. 
Then, the link prediction components from $GCN_{link}$ are used to restore the order of text segments that are grouped into $ConnectedSegments$ based on their linkage relationships. 

For shape approximation (SAp), we first draw the text segments on a blank image and then use the contour of these overlapping text segments to approximate the original text contour. To connect text segments with each other more closely, an opening operation with a kernel size of $3\times 3$ is performed. Similar to~\cite{11,21}, the GGTR map is used %indirectly 
to further suppress the text instances with IoUs less than 0.5.

It is worth noting that our inference method discards the route-finding process of existing bottom-up methods for determining the visiting order of text contour points. Instead, we generate dense text segments to approximate the text contour more accurately with only one NMS processing step.

\begin{figure*}[t]  
\subfigure{
\begin{minipage}[t]{0.18\linewidth}
  
\includegraphics[width=3.4cm,height=3cm]{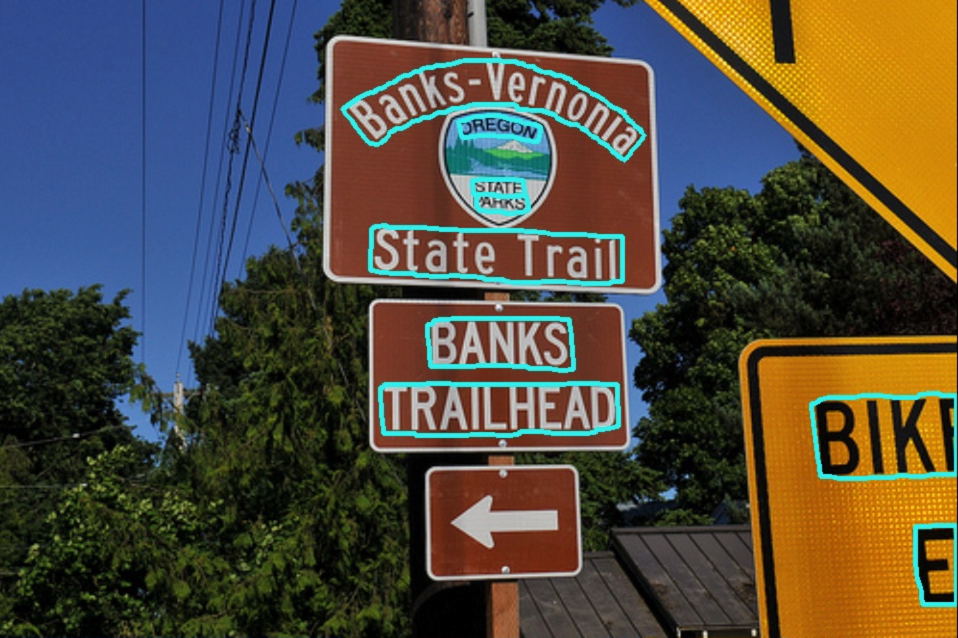}
\label{2a}
\end{minipage}
}               
\subfigure{  
\begin{minipage}[t]{0.18\linewidth}

\includegraphics[width=3.4cm,height=3cm]{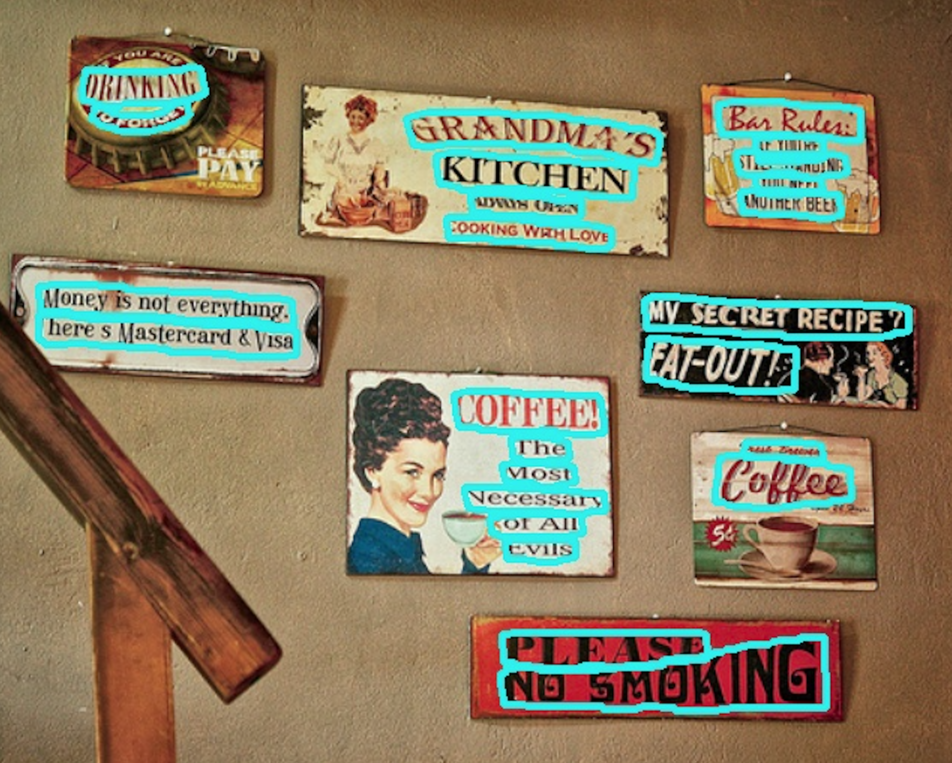}
\label{2b}
\end{minipage}   
} 
\subfigure{  
\begin{minipage}[t]{0.18\linewidth}

\includegraphics[width=3.4cm,height=3cm]{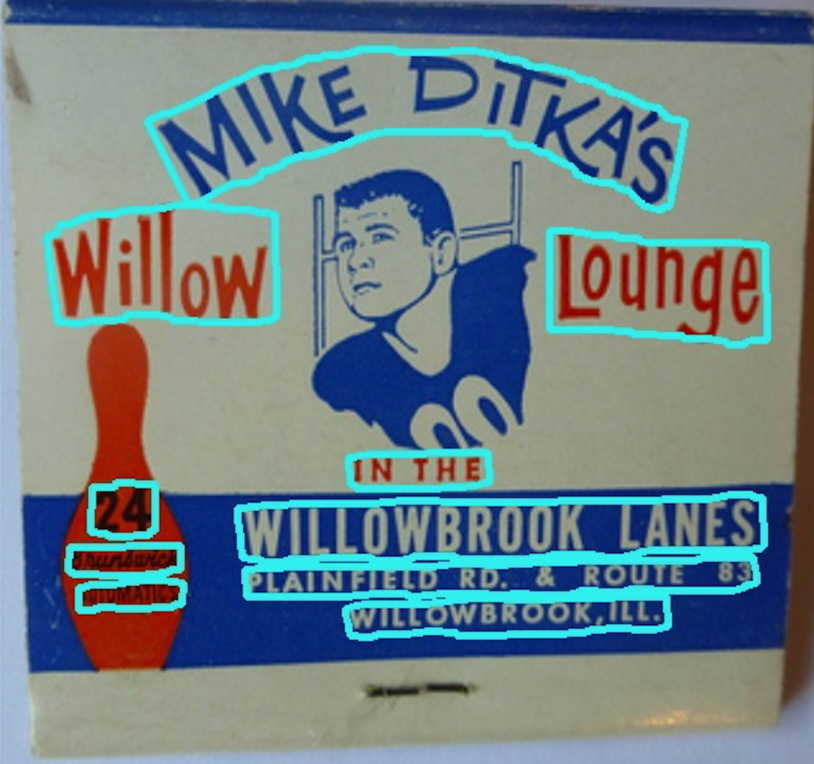}
\label{2b}
\end{minipage}   
} 
\subfigure{  
\begin{minipage}[t]{0.18\linewidth}

\includegraphics[width=3.4cm,height=3cm]{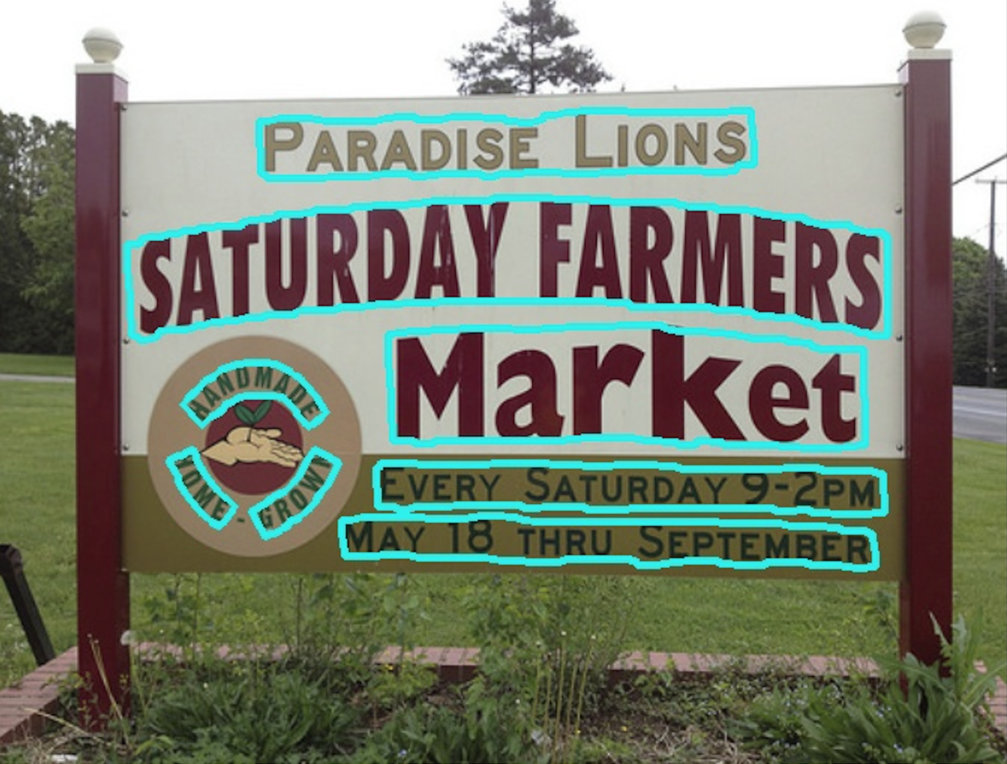}
\label{2b}
\end{minipage}   
} 
\subfigure{  
\begin{minipage}[t]{0.18\linewidth}

\includegraphics[width=3.4cm,height=3cm]{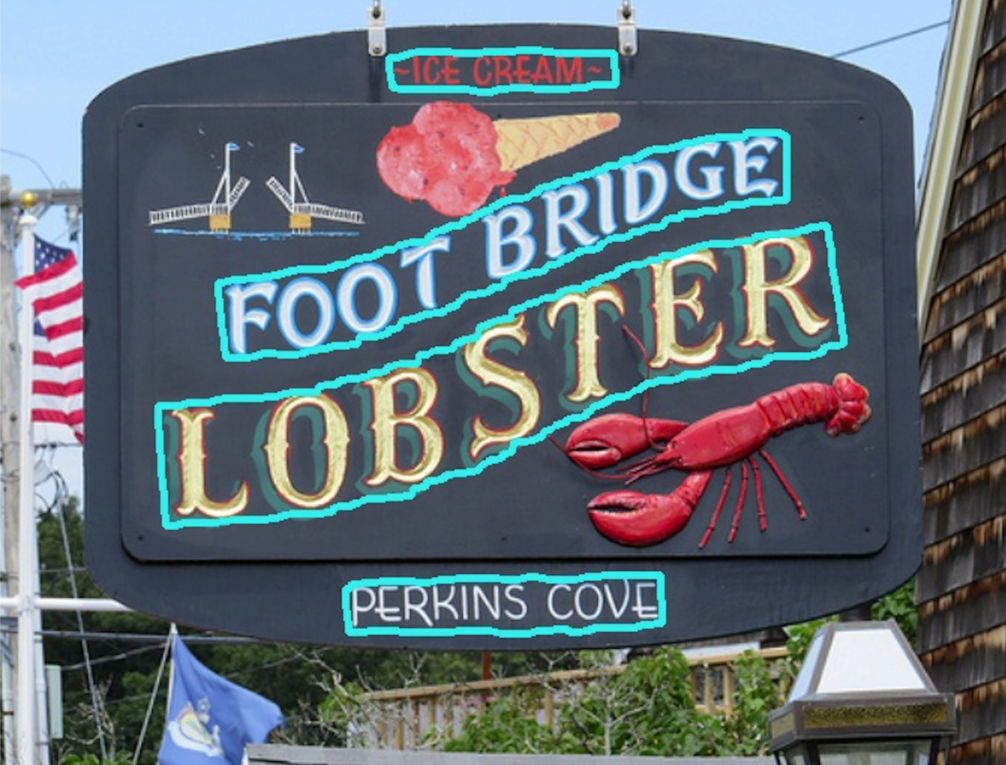}
\label{2b}
\end{minipage}   
} 
%%%%%%%%%%%%%%%%
\vskip -12pt
\subfigure{
\begin{minipage}[t]{0.18\linewidth}
  
\includegraphics[width=3.4cm,height=3cm]{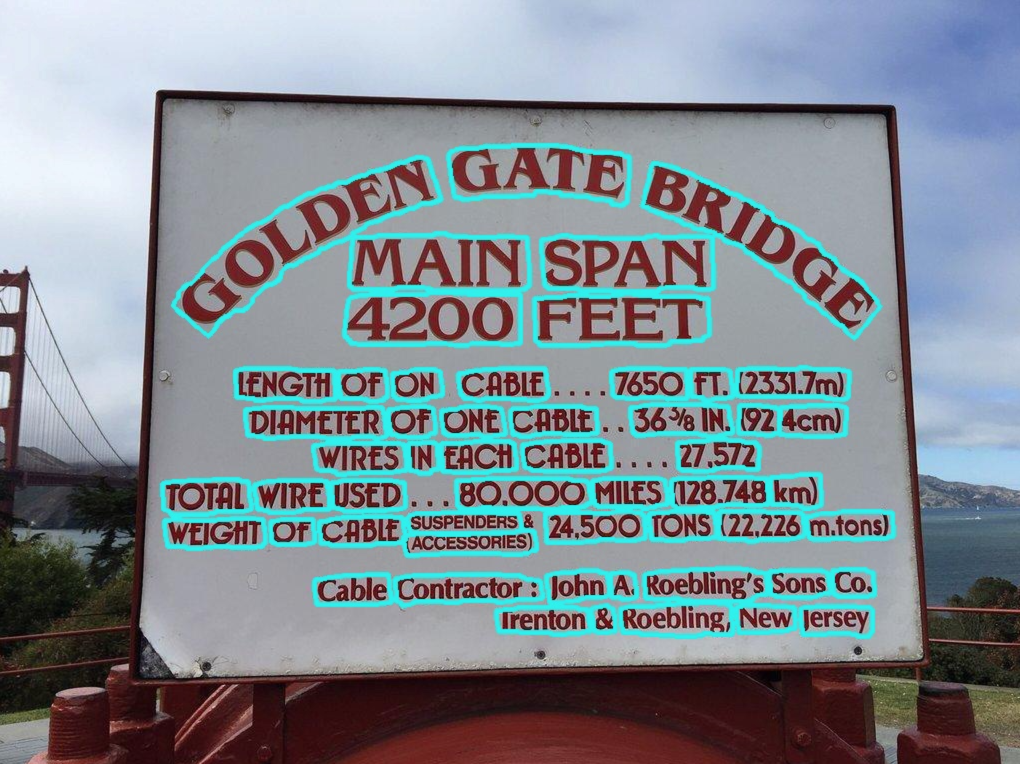}
\label{2a}
\end{minipage}
}               
\subfigure{  
\begin{minipage}[t]{0.18\linewidth}

\includegraphics[width=3.4cm,height=3cm]{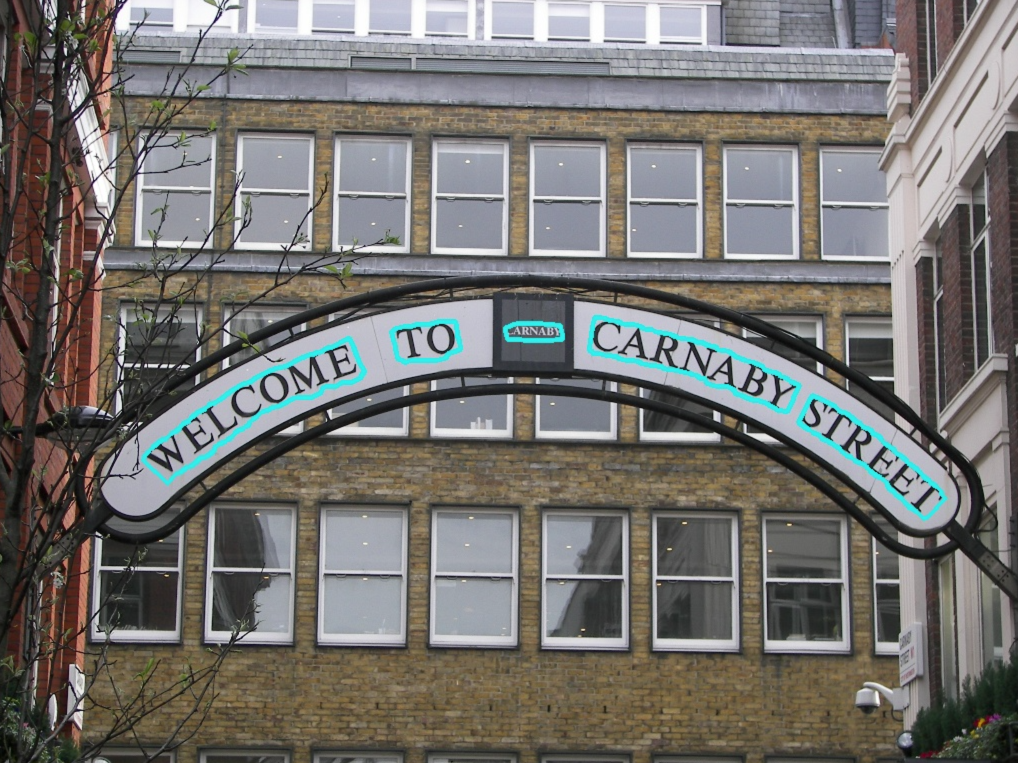}
\label{2b}
\end{minipage}   
} 
\subfigure{  
\begin{minipage}[t]{0.18\linewidth}

\includegraphics[width=3.4cm,height=3cm]{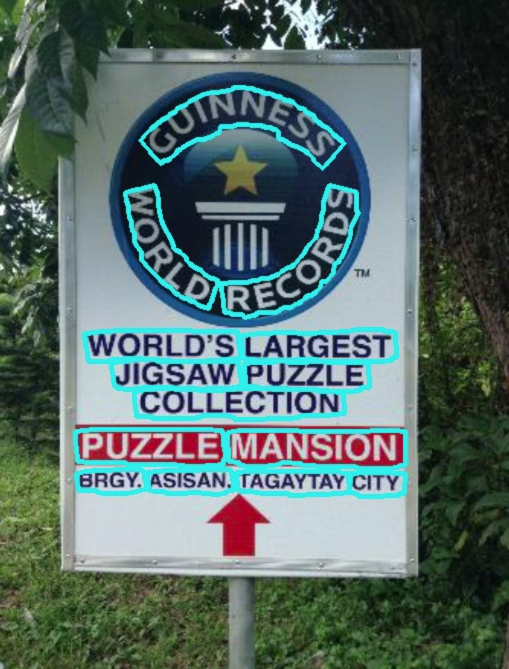}
\label{2b}
\end{minipage}   
} 
\subfigure{  
\begin{minipage}[t]{0.18\linewidth}

\includegraphics[width=3.4cm,height=3cm]{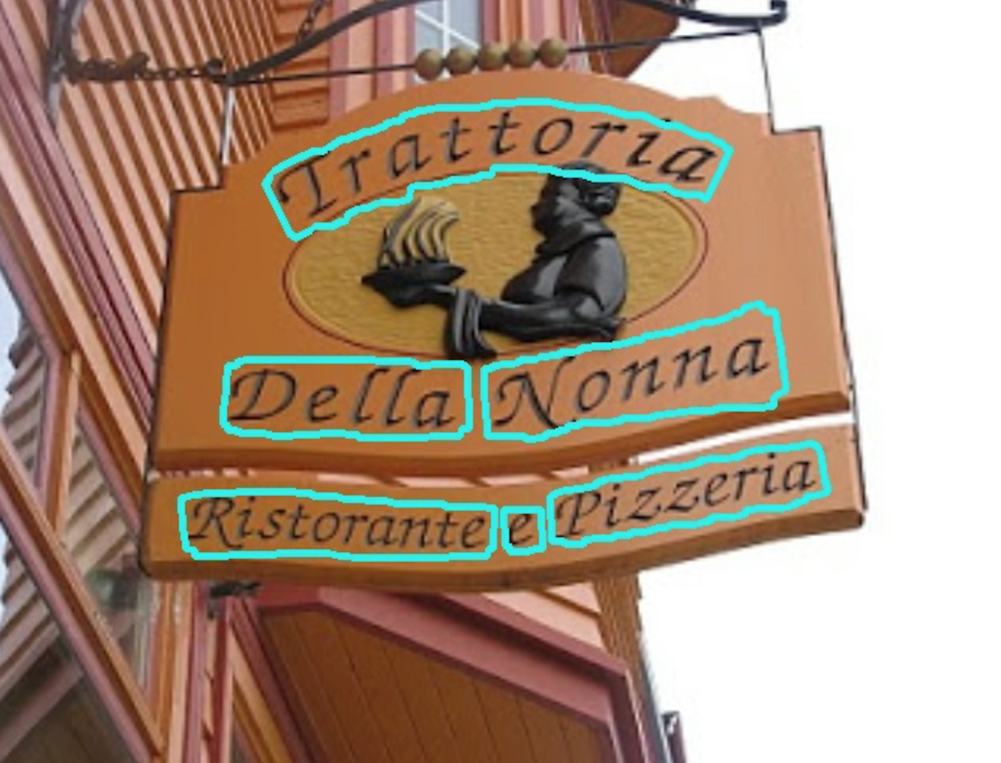}
\label{2b}
\end{minipage}   
} 
\subfigure{  
\begin{minipage}[t]{0.18\linewidth}

\includegraphics[width=3.4cm,height=3cm]{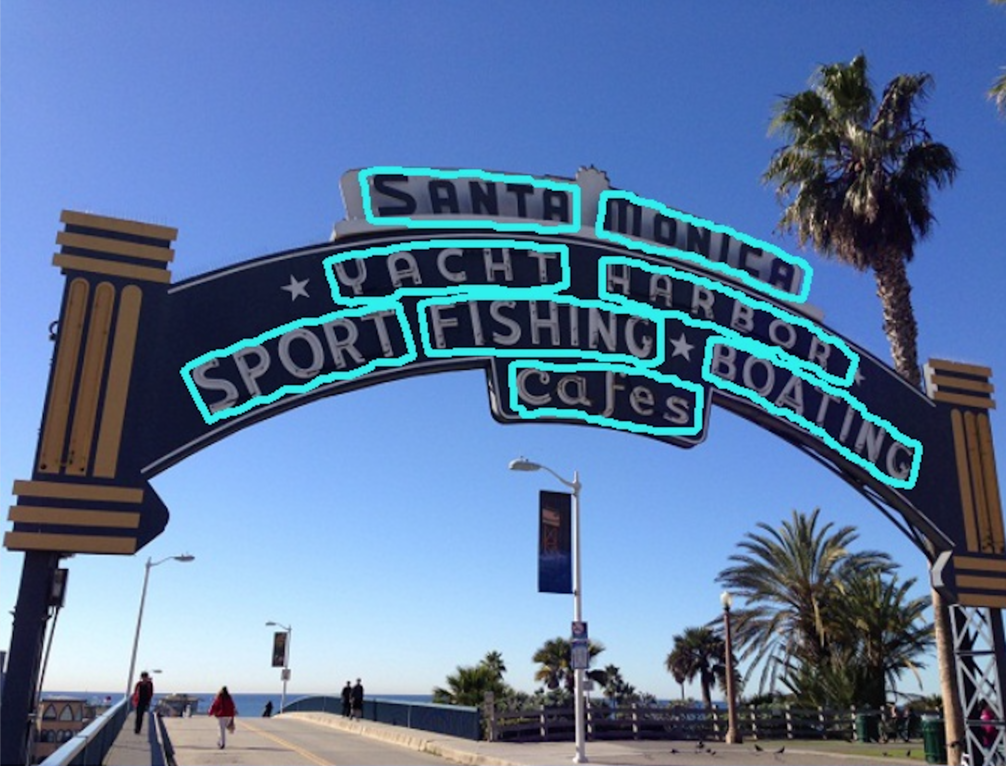}
\label{2b}
\end{minipage}   
} \vskip -12pt
%%%%%%%%%%%%%
\subfigure{
\begin{minipage}[t]{0.18\linewidth}
  
\includegraphics[width=3.4cm,height=3cm]{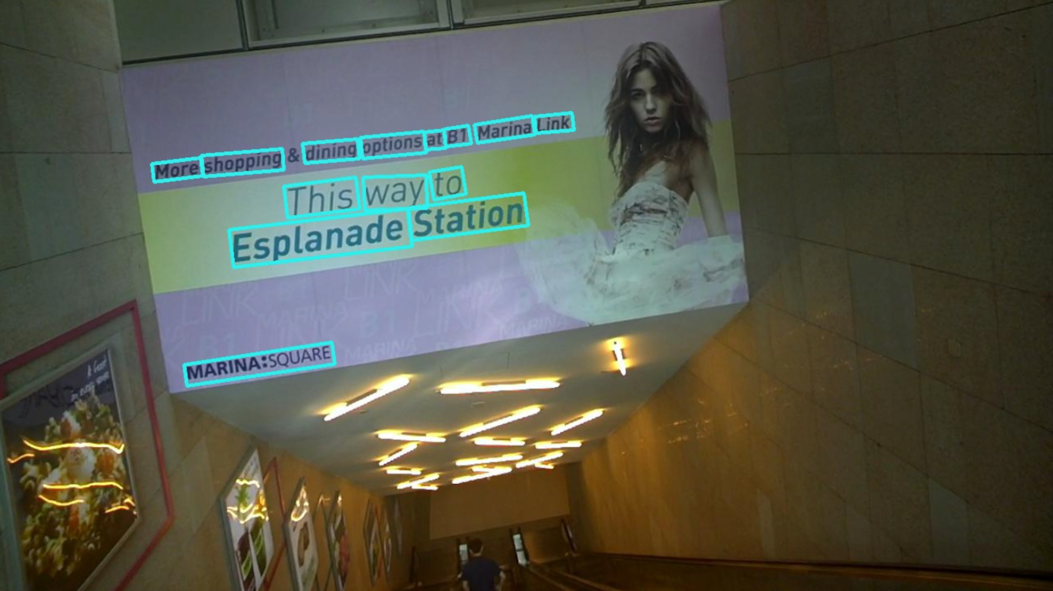}
\label{2a}
\end{minipage}
}               
\subfigure{  
\begin{minipage}[t]{0.18\linewidth}

\includegraphics[width=3.4cm,height=3cm]{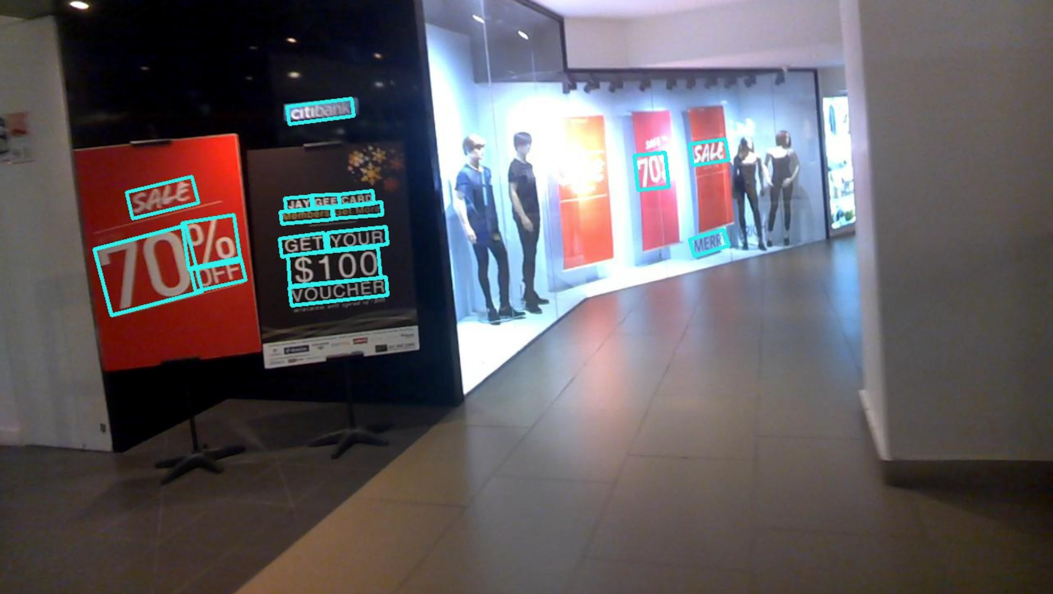}
\label{2b}
\end{minipage}   
} 
\subfigure{  
\begin{minipage}[t]{0.18\linewidth}

\includegraphics[width=3.4cm,height=3cm]{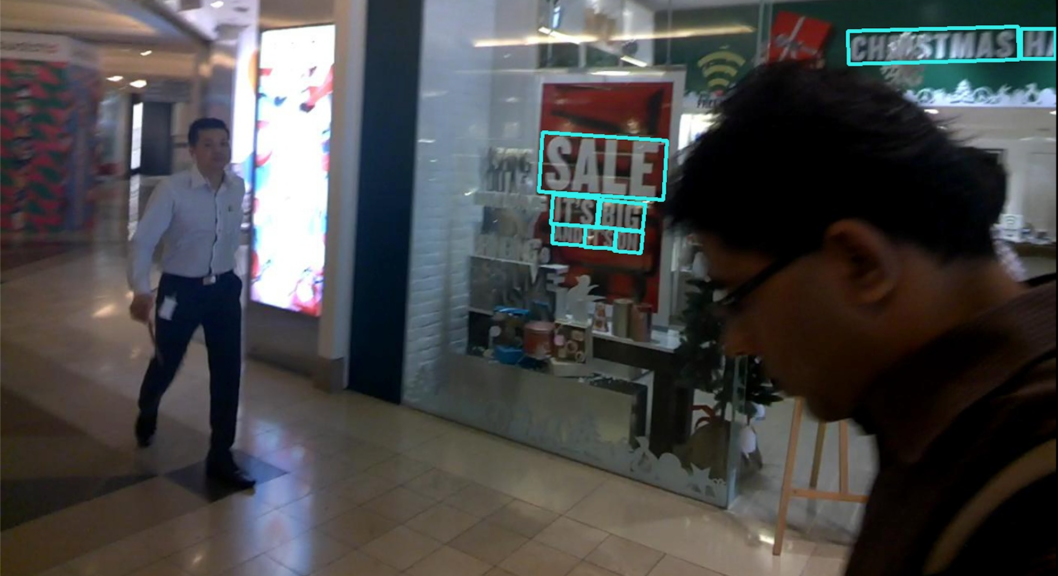}
\label{2b}
\end{minipage}   
} 
\subfigure{  
\begin{minipage}[t]{0.18\linewidth}

\includegraphics[width=3.4cm,height=3cm]{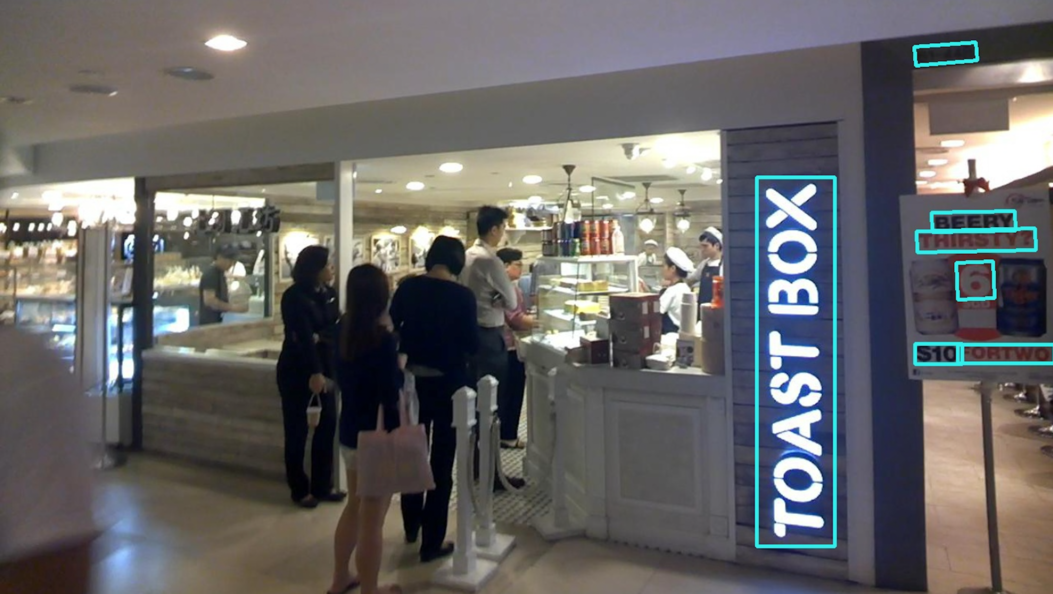}
\label{2b}
\end{minipage}   
} 
\subfigure{  
\begin{minipage}[t]{0.18\linewidth}

\includegraphics[width=3.4cm,height=3cm]{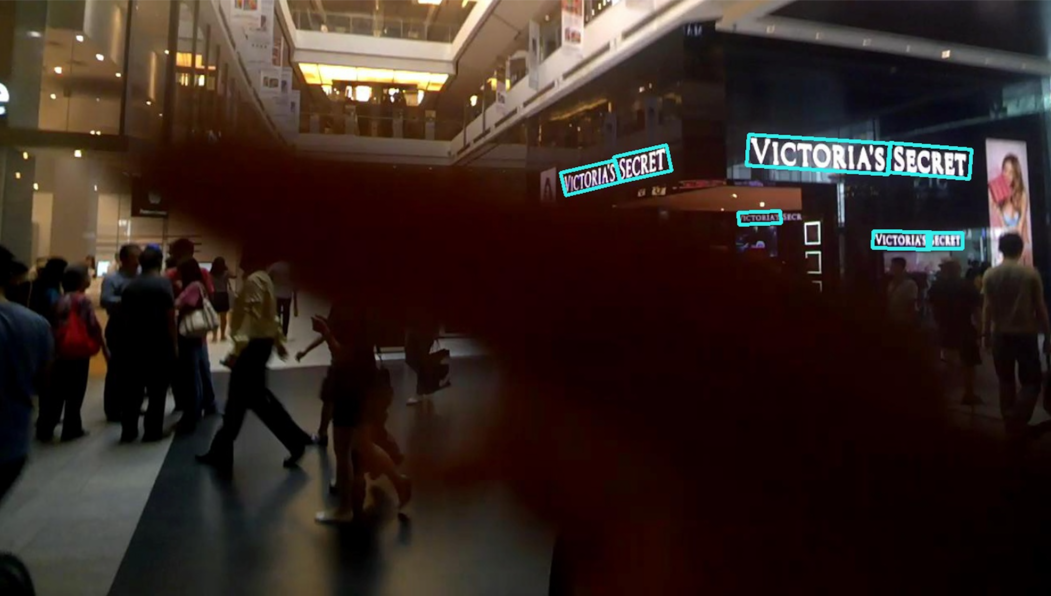}
\label{2b}
\end{minipage}   
} \vskip -12pt
%%%%%%%%%%%%%5
\subfigure{
\begin{minipage}[t]{0.18\linewidth}
  
\includegraphics[width=3.4cm,height=3cm]{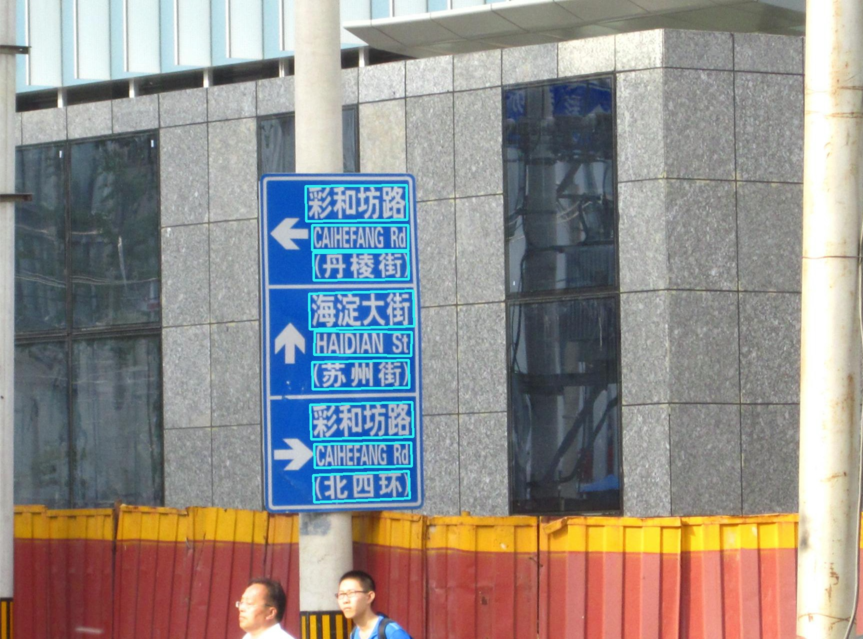}
\label{2a}
\end{minipage}
}               
\subfigure{  
\begin{minipage}[t]{0.18\linewidth}

\includegraphics[width=3.4cm,height=3cm]{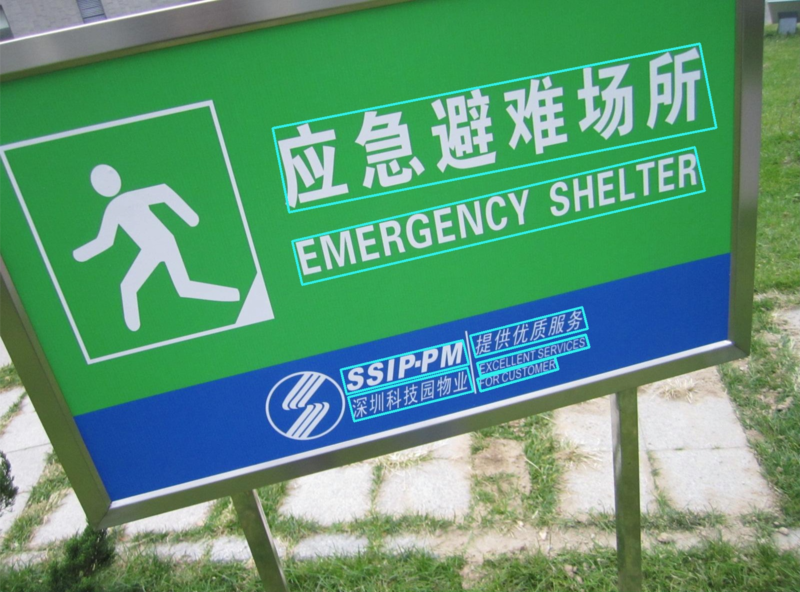}
\label{2b}
\end{minipage}   
} 
\subfigure{  
\begin{minipage}[t]{0.18\linewidth}

\includegraphics[width=3.4cm,height=3cm]{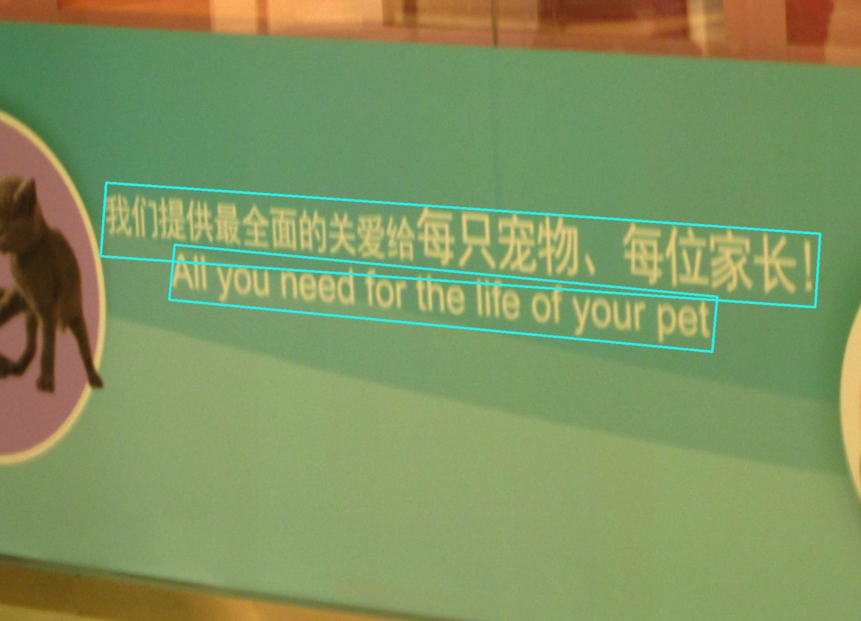}
\label{2b}
\end{minipage}   
} 
\subfigure{  
\begin{minipage}[t]{0.18\linewidth}

\includegraphics[width=3.4cm,height=3cm]{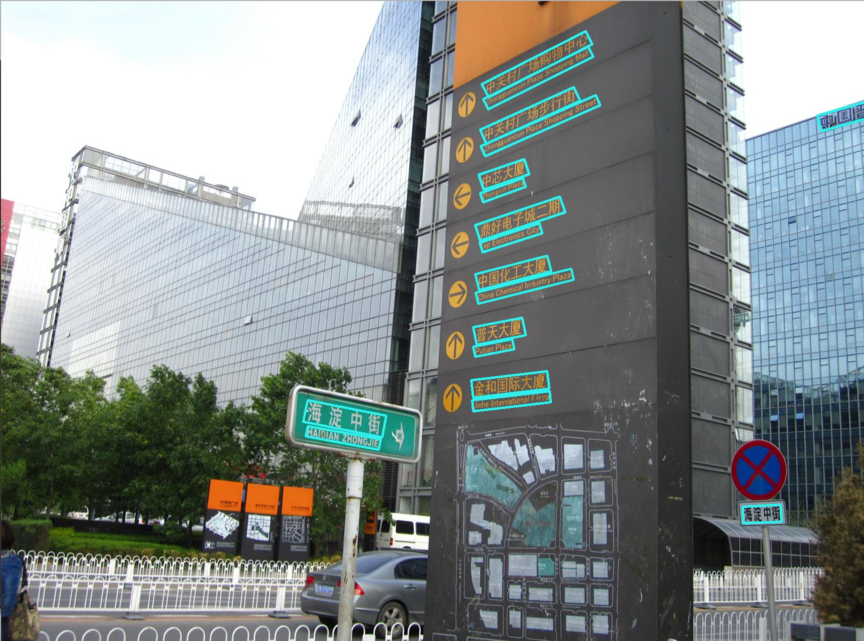}
\label{2b}
\end{minipage}   
} 
\subfigure{  
\begin{minipage}[t]{0.18\linewidth}

\includegraphics[width=3.4cm,height=3cm]{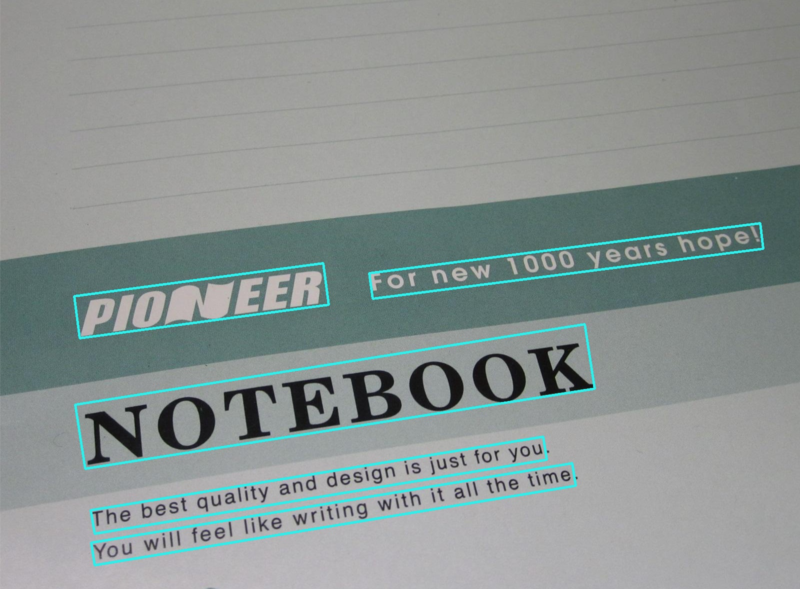}
\label{2b}
\end{minipage}   
} 

%%%%%%%%%%
%\vskip -10pt
\centering
\caption{\add{Visualization of the text detection results obtained on CTW1500 (1st row), Total-Text (2nd row), ICDAR2015 (3rd row) and MSRA-TD500 (4th row).}}
\label{fig:visualresults}
\end{figure*}

%\section{EXPERIMENTS}
\section {Experiments}
\label {sect:exp}

To evaluate the performance of our proposed method, we conducted comprehensive experiments on five mainstream scene text detection datasets, including the curved text datasets Total-Text and CTW1500, the multi-oriented dataset ICDAR2015, and the multi-lingual dataset MLT2017. 
For a fair comparison, VGG16 and ResNet50 were used as the feature extraction backbone to exclude the performance gain brought by the backbone. 
Moreover, only single-scale testing results were compared to exclude the performance gain contributed by multi-scale testing.

\subsection{Datasets}

\noindent {\bfseries SynthText}~\cite{synthetic} consists of 800k synthetic images of multi-oriented text with character-level annotation. %The character-level 
This annotation is used to produce the ground truth of text segment types.\\
{\bfseries Total-Text}~\cite{18} consists of horizontal, multi-oriented, and curved text instances with polygon and word-level annotations. It contains 1,255 training images and 300 testing images.\\
{\bfseries CTW1500}~\cite{19} consists of curved and multi-oriented texts, all annotated by polygons, and tends to label long curved text. It has 1,500 training and 1,000 testing images.\\  
{\bfseries ICDAR2015}~\cite{icdar2015} includes multi-orientated and small-scale text instances. Its ground truth is annotated with word-level quadrangles. It contains 1,000 training and 500 testing images. \\
{\bfseries MSRA-TD500}~\cite{TD500} is dedicated to detecting multi-oriented long non-Latin texts. It contains 300 training images and 200 testing images with word-level annotation. Here, we follow the previous methods~\cite{39,7} and add 400 training images from TR400~\cite{TR400} to extend this dataset.\\
{\bfseries MLT2017}~\cite{MLT} is designed for detecting multi-lingual texts in nine languages with word-level annotation. It contains 9,000 training images and 9,000 testing images.

\subsection{Implementation Details}
Our algorithm is implemented using PyTorch 1.7. The VGG16 is pre-trained with ImageNet, with FPN adopted for multi-scale feature extraction. We conducted our experiments on an RTX3090 GPU with 24GB memory. 
All images used for training and testing are of a single scale. 
For training, the images are resized to $640\times 640$ for images from CTW1500 and Total-Text, and $1280 \times 1280$ for images from ICDAR2015, MSRA-TD500 and MLT2017. 
Data augmentation techniques including rotation, random cropping, color variations, adding random noise, blur, and changing lightness were adopted. The batch size was set to 10.

We first used SynthText to pre-train our model for 10 epochs using Adam optimizer and a learning rate of $0.001$. 
Then, we fine-tuned our model on real text benchmark datasets for 800 epochs with SGD optimizer and a learning rate of $0.001$. The momentum of SGD was set to $0.9$. For testing, the short side of images was kept at $640$ pixels for CTW1500 and Total-Text, and 1,280 pixels for ICDAR2015, MSRA-TD500 and MLT2017, while retaining their aspect ratios.

\subsection{Comparison with State-of-the-art Methods}

Examples of the text detection results obtained with our proposed method are shown in Fig.~\ref{fig:visualresults}. 
Table~\ref{tab:all} shows quantitative comparisons of the detection results with the state-of-the-art approaches on all mainstream benchmark datasets, \textit{i.e.}, CTW1500, Total-Text, ICDAR2015, MSRA-TD500 and MLT2017, respectively. 
The ``-" in the table indicates that the comparative method did not report results on the dataset.

\begin{figure*}[h]
	%\scriptsize
	\renewcommand{\tabcolsep}{1pt} % adjust horizontal space
	\renewcommand{\arraystretch}{0.8} % adjust vertical space
	\centering
	\begin{tabular}{cccc}
	    \includegraphics[width=4cm,height=3cm]{6fuse.jpg} &
	    \includegraphics[width=4cm,height=3cm]{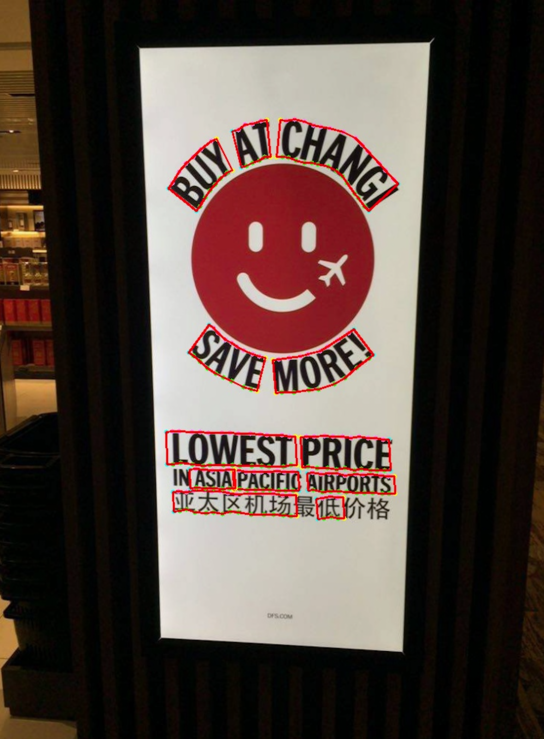} &
	    \includegraphics[width=4cm,height=3cm]{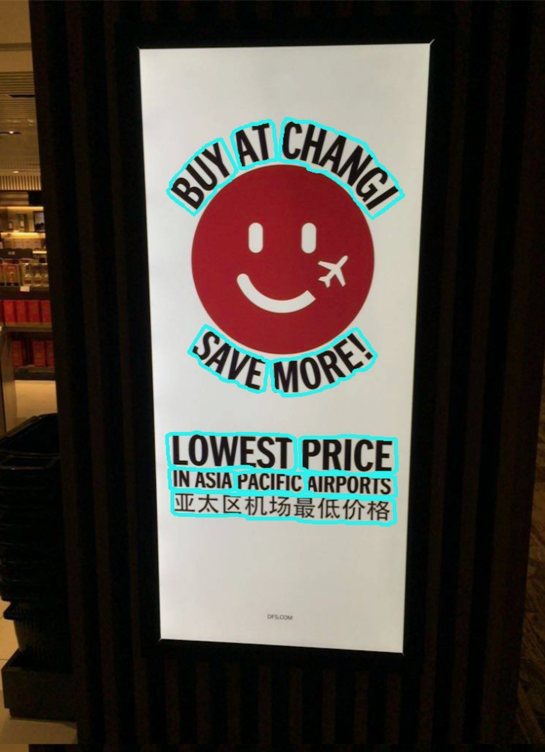} &
	    \includegraphics[width=4cm,height=3cm]{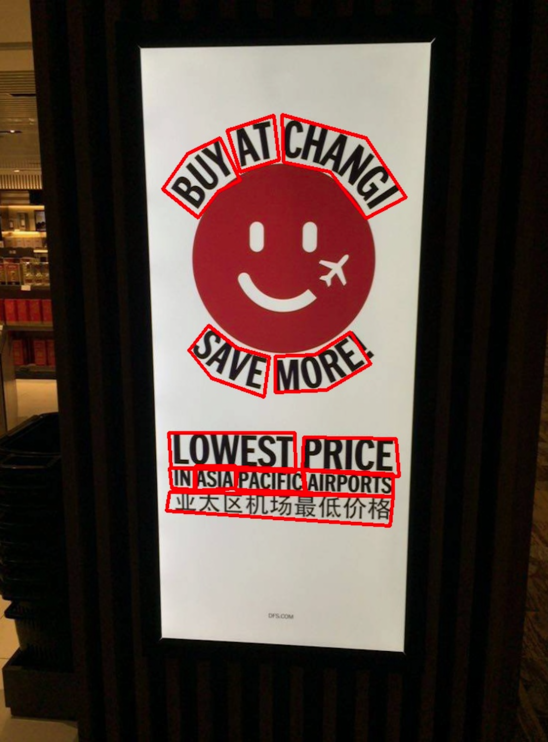}
		\\
	    \includegraphics[width=4cm,height=3cm]{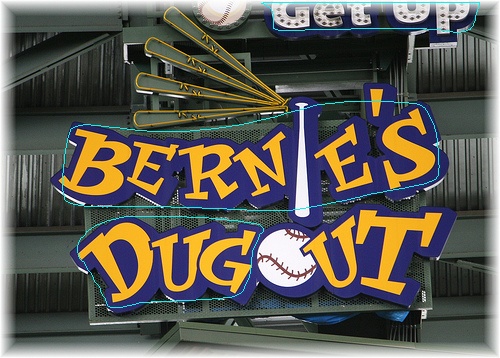} &
	    \includegraphics[width=4cm,height=3cm]{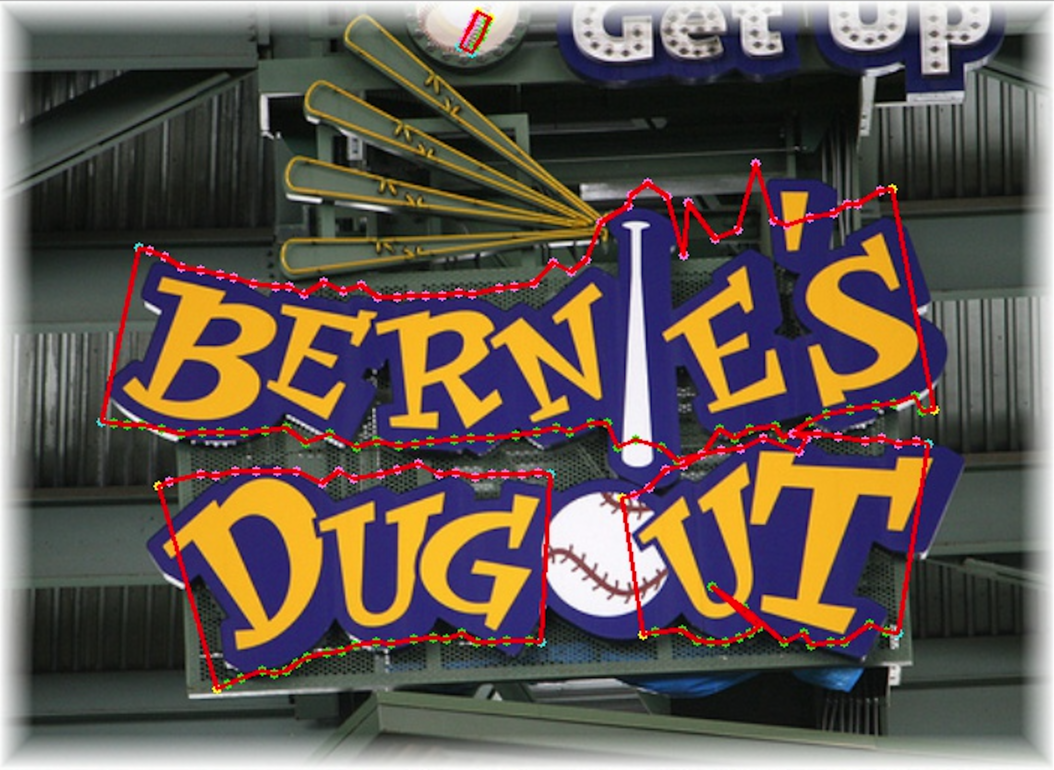} &
	    \includegraphics[width=4cm,height=3cm]{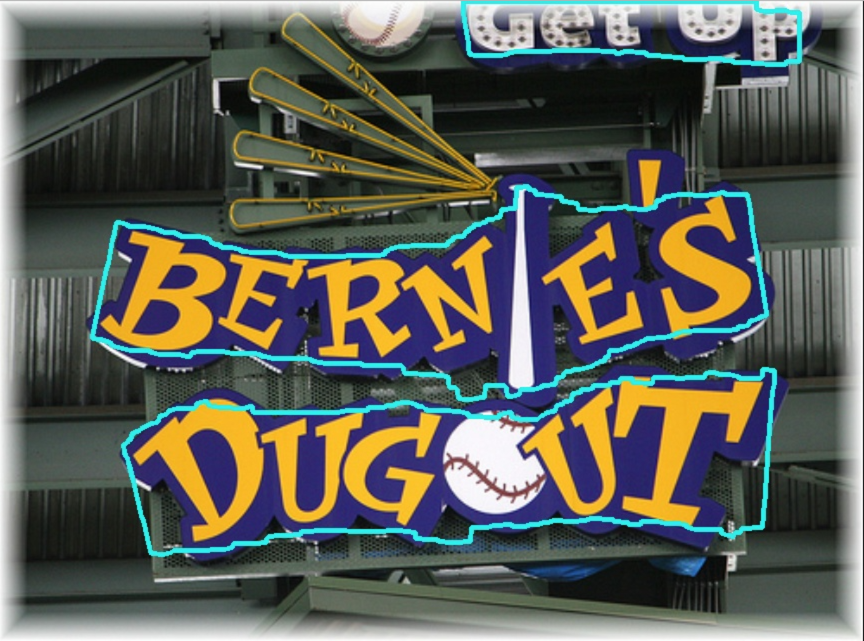} &
	    \includegraphics[width=4cm,height=3cm]{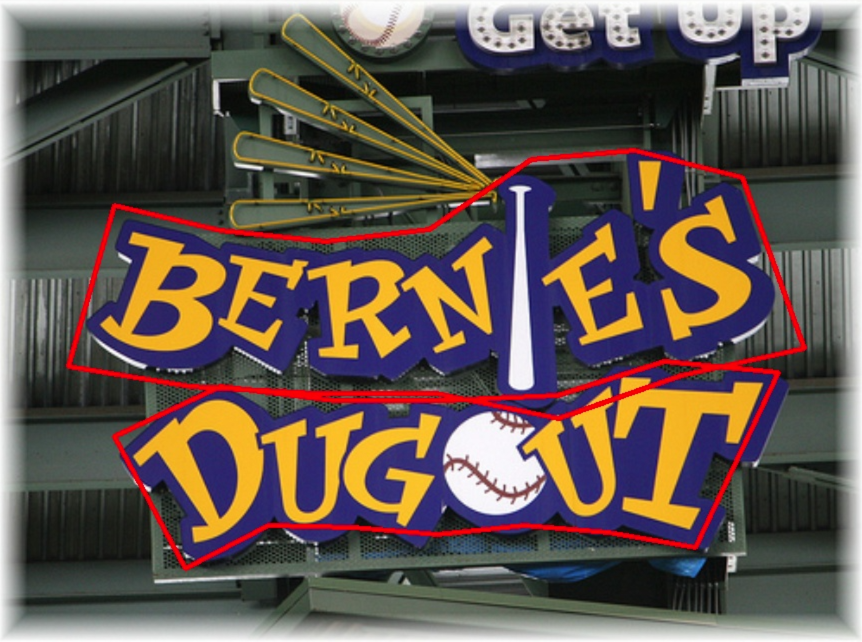}
		\\

		\\
		(a) Top-down method~\cite{9} & (b) Bottom-up method~\cite{11} & (c) Ours & (d) Ground truth
		\\
	\end{tabular}
	\vspace{-1mm}
	\caption{\add{Qualitative comparisons with the SOTA top-down~\cite{9} and bottom-up~\cite{11} methods on some challenging examples.}}
	\label{fig:compare drrg}
\end{figure*}

{\bfseries Curved text detection:} 
Our results on CTW1500 and Total-Text, the two most representative benchmark datasets of arbitrary shape text, demonstrate the effectiveness of our algorithm in dealing with the highly curved and spatially separated texts. 
As shown in Table~\ref{tab:all}, our method has achieved F-measure scores of $86.4\%$ and $87.6\%$ on these two datasets, respectively, which outperforms all existing state-of-the-art (SOTA) methods.   
Specifically, our method remedies the shortcomings of the bottom-up methods~\cite{8,11,7} 
and can achieve SOTA results with only a VGG16 backbone without introducing deformable convolutional blocks to increase the backbone network's feature extraction ability as in~\cite{7,39,43,tmm3,tmm1}. 
Compared with the SOTA top-down methods~\cite{36,43}, our method shows improvements of 0.9\% on CTW1500 and 0.7\% on Total-Text respectively, thanks to our proposed false-detection suppression and shape-approximation strategies, as well as the dense overlapping design of text segments.  
Different from~\cite{9} requiring a weakly supervised instance segmentation to predict the mask and class of each character, our method focuses on classifying the types of text segments without extra segmentation. 
\add{Fig.~\ref{fig:compare drrg} qualitatively compares the visual detection results of our proposed method with results obtained by the GCN-based bottom-up approach DRRG~\cite{11} and Mask RCNN based top-down method TextFuseNet~\cite{9} on some challenging text instances in CTW1500 and Total-Text. 
As shown in this comparison, 
our proposed false-detection suppression and shape-approximation strategies are effective in handling false detection and the route-finding failure.}

The SOTA results obtained on these two datasets support our claim that the relative underachievements of existing bottom-up methods are not caused by the limited feature capturing ability of the text proposal backbones or GCN. 
Also, bottom-up methods can be superior to top-down methods when adopting our proposed false-positive/negative suppression and shape-approximation strategies in the inference stage.

\begin{table*}[t]
  \caption{Results on CTW1500, Total-Text, ICDAR2015, MSRA-TD500 and MLT2017. (\add{P: Precision, R: Recall, F: F-measure}, \dag: bottom-up, \S: top-down, *: GCN based methods)}
  \label{tab:all}
  \setlength\tabcolsep{2pt}
   \centering
  \begin{tabular}{c|c|c|ccc|ccc|ccc|ccc|ccc}
    \toprule
    \hline
      \multirow{2}{*}{Method}& \multirow{2}{*}{Venue} &  \multirow{2}{*}{Backbone} & \multicolumn{3}{c|}{CTW1500} &  \multicolumn{3}{c|}{Total-Text}  &  \multicolumn{3}{c|}{ICDAR2015} &  \multicolumn{3}{c|}{MSRA-TD500}&  \multicolumn{3}{c}{MLT2017}\\ 
		\cline{4-18}
		 &&&P(\%) & R(\%) & F(\%) & P(\%) & R(\%) & F(\%)& P(\%) & R(\%) & F(\%)& P(\%) & R(\%) & F(\%)& P(\%) & R(\%) & F(\%) \\
    \hline
    TextSnake\dag\cite{21} &ECCV'18&VGG16  &67.9 &85.3 & 75.6    &82.7 &74.5& 78.4    &84.9 &80.4 &82.6      &83.2 &73.9& 78.3   &- &- &-\\
    LOMO\S \cite{16}      &CVPR'19&ResNet50&85.7 &76.5&80.8     &88.6 &75.7 &81.6    &91.3 &83.5 &87.2       &- &- &-   &78.8 &60.6 &68.5\\
    TextRay\dag \cite{12}    &MM'20&ResNet50&82.8 &80.4 &81.6  &83.5 &77.8 &80.6    &- &- &-                &- &- &-  &- &- &-\\
    OPOM\dag \cite{tmm3} &TMM'20&ResNet50&85.1&80.8&82.9   &88.5&82.9&85.6  &89.1&85.5&87.3  &86.0&83.4&84.7  &82.9&70.5&76.2\\
    DB\S \cite{39}          &AAAI'20&ResNet50& 86.9 &80.2 &83.4   &87.1 &82.5 &84.7    & 91.8 &83.2 &87.3      & {\bfseries91.5} &79.2 &84.9 & 83.1 &67.9 &74.7\\
    CRAFT\dag \cite{6}       &CVPR'19&VGG16& 86.0 &81.1 &83.5     &86.0& 81.1& 83.5    & 89.8 &84.3& 86.9      & 88.2& 78.2& 82.9    & 80.6& 68.2& 73.9\\
    TextDragon\dag \cite{29}  &ICCV'19 &VGG16& 84.5 &82.8 &83.6   &85.6 &75.7 &80.3    & {\bfseries92.5} &83.8 &87.9 & - &- &-   & - &- &-\\
    SDM\S \cite{36} &ECCV'20 &ResNet50& 85.8 &82.2 &84.0   &89.3 &84.7 &86.9 & 88.7 &88.4 &88.5& - &- &- & {\bfseries84.2} &72.8 &78.1\\
    PSE\dag \cite{PSE} &CVPR'19 &ResNet50& 84.8 &79.7 &82.2&84.0&79.0&80.1   &86.9 &84.5 &85.7 & - &- &-  & 73.8 &68.2 &70.9\\
    PAN\S \cite{PAN} &ICCV'19 &ResNet50& 86.4 &81.2 &83.7   &89.3 &81.0 &85.0 & 84.0 &81.9 &82.9 & 84.4 &83.8 &84.1 & - &- &-\\
    *PuzzleNet\dag \cite{8}    &arXiv'20&ResNet50& 84.1 &84.7 &84.4 & - &-&-            &  89.1 &86.9 &88.0     & 88.2 &83.5&85.8  & - &-&-\\
    *DRRG\dag \cite{11}          &CVPR'20&VGG16& 85.9 &83.0 &84.4 & 86.5& 84.9 &85.7   & 88.5 &84.7 &86.6      &88.1& 82.3 &85.1 &74.5& 61.0 &67.3\\
     PCR \S \cite{42}         &CVPR'21&DLA34~\cite{dla}& 87.2 &82.3&84.7& 88.5 &82.0&85.2& - &-&-& 90.8
    &83.5&{\bfseries87.0}&-&-&-\\
     Dai et al.\S\cite{tmm1}    &TMM'21&ResNet50&86.2&80.4&83.2  &85.4&81.2&83.2 &87.2&81.3&84.1 &-&-&- &-&-&-\\
    *ReLaText\dag \cite{7}    &PR'20 &ResNet50& 86.2&83.3 &84.8 &84.8& 83.1 &84.0    &-& - &-                &90.5& 83.2 &86.7\\
    CountourNet\S \cite{10}   &CVPR'20&ResNet50& 85.7 &84.0 &84.8 &86.9& 83.9 &85.4    & 86.1 &87.6 &86.9      & -& - &-   &-&-&-\\
    Dai et al.\S \cite{tmm2}       &TMM'21&ResNet50&85.7&85.1&85.4 &84.6&78.6&81.5 &86.2&82.7&84.4 &-&-&- &79.5&66.8&72.6\\
    FCENet\S \cite{43}         &CVPR'21&ResNet50& 87.6 &83.4&85.5  & 89.3 &82.5&85.8 & 90.1 &82.6&86.2 &-&-&- &-&-&-\\
    TextFuseNet\S\cite{9}    &IJCAI'20 &ResNet50& 85.0 &{\bfseries85.8} &85.4   &87.5& 83.2 &85.3 & 88.9 &{\bfseries91.3} &{\bfseries90.1} & - & - &-  &-&-&-\\

    \hline

    *Ours\dag   &-     &VGG16& {\bfseries89.8} &83.3 &{\bfseries86.4} & {\bfseries89.4} &{\bfseries85.8} &{\bfseries87.6}& 91.4 &87.7 &89.5& 90.5 &{\bfseries83.8} &{\bfseries87.0} & 82.9 &{\bfseries74.0} &{\bfseries78.2}\\
  \hline
  
\end{tabular}
\end{table*}

{\bfseries Multi-oriented text detection:} 
On the dataset ICDAR2015, as shown in Table~\ref{tab:all}, our method has also achieved a comparable SOTA result of an F-measure of $89.5\%$, outperforming some high-performing top-down methods such as FCENet~\cite{43} and ContourNet~\cite{10} by more than 2.5 percentage points. 

Our result is slightly lower than that of the TextFuseNet method~\cite{9}. 
We argue that TextFuseNet adopted an instance segmentation strategy, which introduced %a
classification (recognition) %step 
to enhance detection results.  
Also, the approaches of~\cite{8,7,9,10} have adopted a multi-scale training/testing strategy, which is a widely-known tip used on ICDAR2015 that can boost the detection accuracy significantly on this dataset, 
whereas we only use single-scale training and testing rather than multi-scale training or testing for a fair comparison. 

On the dataset MSRA-TD500, as shown in Table~\ref{tab:all}, our method has equalled the SOTA result with an F-measure of $87.0\%$. 
Moreover, thanks to our proposed FPNS strategy, our method has achieved the highest recall rate of $83.8\%$. This shows the effectiveness of our idea of depicting the streamline of the texts, which helps to retrieve some misdetected texts.  
As our approach focuses on detecting arbitrary-shape text, it is a common situation that arbitrary-shape text detectors cannot significantly improve the SOTA results in both multi-oriented detection benchmarks, especially ICDAR2015, which exhibits large scale variations of texts. 
For example, the arbitrary-shape text detectors~\cite{tmm3,8,11,7,10,tmm2,43,tmm1,42} can only achieve SOTA results on at most one multi-oriented detection benchmark.
The arbitrary-shape text detectors focus more on the space variation of texts, whereas multi-oriented text detectors focus more on scale variation.

{\bfseries Multi-lingual text detection:} 
As shown in Table~\ref{tab:all}, on the multi-lingual scene text dataset MLT2017, our network has also surpassed the SOTA method SDM~\cite{36} with a Precision of $82.9\%$, a Recall of $74.0\%$ and an F-measure of $78.2\%$. 
This is because the dense design of the text segments can effectively depict the `characterness' property that non-Latin texts also exhibit. 
Moreover, our proposed FPNS and SAp strategies enable the network to accurately identify connectivity within multi-lingual texts. 
The highest F-measure demonstrates that our method has good stability in detecting multi-lingual texts. 

\begin{table}[h]
  \caption{The impact of our proposed FPNS and SAp strategies.}
  %\footnotesize
  \label{tab:1}
   \centering
    \setlength\tabcolsep{5pt}
  \begin{tabular}{c|c|c|c|c|ccc}
    \toprule
    \hline
    Datasets& GCN & Node& GGTR &SAp &P (\%) &R (\%) &F (\%)\\
    \hline
    $\quad$&$\checkmark$&$\times$&$\times$&$\times$&85.9 &83.0& 84.4\\
  
    $\quad$&$\checkmark$&$\checkmark$&$\times$&$\times$&87.6 &82.3& 84.8\\
  
    CTW1500&$\checkmark$&$\times$&$\checkmark$&$\times$&87.4 &82.6& 84.9\\
  
    $\quad$&$\checkmark$&$\times$&$\times$&$\checkmark$&87.2 &83.1& 85.1\\
  
     $\quad$&$\checkmark$&$\checkmark$&$\checkmark$&$\times$&88.6 &{\bfseries83.4}& 85.9\\
    $\quad$&$\times$&$\times$&$\checkmark$&$\checkmark$&88.6 &82.6& 85.5\\
  
    $\quad$&$\checkmark$&$\checkmark$&$\checkmark$&$\checkmark$&{\bfseries89.8} &83.3 &{\bfseries86.4}\\

    %\hline
    \hline
    %\midrule
    $\quad$&$\checkmark$&$\times$&$\times$&$\times$&86.5 &84.9& 85.7\\
    
    $\quad$&$\checkmark$&$\checkmark$&$\times$&$\times$&86.8 &85.1& 85.9\\
    
    Total-Text&$\checkmark$&$\times$&$\checkmark$&$\times$&87.7 &84.7& 86.2\\
    
        $\quad$&$\checkmark$&$\times$&$\times$&$\checkmark$&87.6 &85.0&86.3\\
        
     $\quad$&$\checkmark$&$\checkmark$&$\checkmark$&$\times$&88.9 &85.0& 86.9\\
     
    $\quad$&$\times$&$\times$&$\checkmark$&$\checkmark$&87.9 &84.8& 86.3\\
    
    $\quad$&$\checkmark$&$\checkmark$&$\checkmark$&$\checkmark$&{\bfseries89.4} &{\bfseries85.8}& {\bfseries87.6}\\
    %\hline
    \hline
    %\midrule
    $\quad$&$\checkmark$&$\times$&$\times$&$\times$&88.7 &86.4& 87.5\\
    
    $\quad$&$\checkmark$&$\checkmark$&$\times$&$\times$&89.0 &86.8& 87.9\\
    
    ICDAR2015&$\checkmark$&$\times$&$\checkmark$&$\times$&88.8 &87.9& 88.3\\
    
        $\quad$&$\checkmark$&$\times$&$\times$&$\checkmark$&89.0 &88.3& 88.6\\
        
     $\quad$&$\checkmark$&$\checkmark$&$\checkmark$&$\times$&90.5 &87.0& 88.7\\
     
    $\quad$&$\times$&$\times$&$\checkmark$&$\checkmark$&91.0 &87.6& 89.3\\
    
    $\quad$&$\checkmark$&$\checkmark$&$\checkmark$&$\checkmark$&{\bfseries91.4} &{\bfseries87.7}& {\bfseries89.5}\\
  %\hline
  \hline
      %\midrule
    $\quad$&$\checkmark$&$\times$&$\times$&$\times$&88.2 &83.0& 85.5\\
    
    $\quad$&$\checkmark$&$\checkmark$&$\times$&$\times$&88.9 &83.1& 85.9\\
    
MSRA&$\checkmark$&$\times$&$\checkmark$&$\times$&89.6 &82.8& 86.1\\
    
        TD500&$\checkmark$&$\times$&$\times$&$\checkmark$&89.0 &83.6& 86.2\\
         
         $\quad$&$\checkmark$&$\checkmark$&$\checkmark$&$\times$&89.8 &83.6& 86.6\\ 
         
     $\quad$&$\times$&$\times$&$\checkmark$&$\checkmark$&89.9 &83.4& 86.5\\
     
    $\quad$&$\checkmark$&$\checkmark$&$\checkmark$&$\checkmark$&{\bfseries90.5} &{\bfseries83.8}& {\bfseries87.0} \\
\hline
\end{tabular}
\end{table}

\subsection{Ablation Studies}
\label{ablation}

\subsubsection {Effectiveness of the Proposed FPNS and SAp Strategies}

To verify the effectiveness of our proposed FPNS and SAp strategies, we conducted ablation studies on the CTW1500, Total-Text, ICDAR2015 and MSRA-TD500 datasets. 

Table~\ref{tab:1} shows the comparison results. Here, we use ``P'', ``R'' and ``F'' to represent Precision, Recall and F-measure, respectively. 
'`Node'' refers to the node-classification-based FPNS strategy and ``GGTR'' refers to the GGTR-based FPNS strategy.
``SAp'' denotes our proposed shape-approximation strategies. 
The baseline is the GCN-based bottom-up method with contour route-finding~\cite{11}.

A few observations can be made from Table~\ref{tab:1}: \\
i) The GCN-Node-classification-based FPNS, \textit{i.e.}, ``FPNS (Node)'', takes weakly-supervised GCN node classification results into account and improved the F-measures by $0.4\%$, $0.2\%$, $0.4\%$ and $0.4\%$ on the CTW1500, Total-Text, ICDAR2015 and MSRA-TD500 datasets, respectively. 
The GGTR based FPNS fuses multi-modal features and improved the F-measures by $0.5\%$, $0.5\%$, $0.8\%$ and $0.6\%$ on the same datasets. 
When both FPNS strategies (\textit{i.e.}, Node classification based and GGTR based FPNS) are adopted, they improved the F-measures by $1.5\%$, $1.2\%$, $1.2\%$ and $1.1\%$ on the tested datasets. 
We argue that the further performance gains are because the objectives of GGTR and node classification well align with each other in terms of effectively removing text-like interference and therefore %the two operations 
they benefit each other.\\
ii) Our proposed shape-approximation strategy ``SAp'' has improved the F-measures by $0.7\%$, $0.6\%$, $1.1\%$ and $0.7\%$ on the tested datasets, because it avoids the problem of the route-finding process being trapped at a local maximum.\\    
iii) Furthermore, when both FPNS and SAp strategies were used together, they achieved larger improvements on all tested datasets by $2.0\%$, $1.9\%$, $2.0\%$ and $1.5\%$, respectively. 
This is because FPNS removes those noisy segments that affect the shape-approximation, resulting in more accurate text regions.

Interpreting the results, the FPNS (GGTR) contains two parts in our work: the first part calculates the union between the GGTR map and the TCL map, resulting in more potential text segment candidates than directly multiplying the two maps containing visual features together (such as in methods~\cite{11,21}), and the second part removes false detections by measuring IoU with the GGTR map. Both %parts 
rectify false detections from a visual-relational perspective, as the GGTR map results from visual-relational fusion.
FPNS (Node) rectifies false detections by measuring attributes of the text segments in local graph structures and upgrades GCNs to a multiple-task network rather than only linkage reasoning, modifications which support each other. Both node classification and link prediction utilize the same relational features and boost each other's performance. This explains why the overall performance is often further improved when both FPNS (Node) and FPNS (GGTR) are applied. The performance improvements reflect that our FPNS strategies can suppress false detections while not overly affecting true detections.

Similarly, the improved recall brought by SAp can be attributed to the proposed SAp using the contour of the dense text segments to approximate the contour of a text instance. This avoids the typical route-finding failure of getting stuck in sub-optimal local maxima that causes misdetection. 
Moreover, when all three strategies are combined, both the SAp's closing operation and FPNS (Node) affect the shape and size of text segment grouping during the FPNS (GGTR) process. 
The results show that our method is 
 effective in terms of suppressing false positives/negatives.

\subsubsection {The Impact of GCN}
We also conducted additional experiments with and without using GCN, while keeping FPNS and shape-approximation.
In this ablation experiment, the TR map was obtained from the original FPN layer, whose weight is shared with the TCL map, without the guidance of GCN.  
The results listed under the heading `GCN' show that even without GCN, our proposed FPNS and SAp strategies can still produce F-measure improvements of $1.1\%$, $0.6\%$, $1.8\%$ and $1.0\%$, respectively.

We argue that GCN makes limited contributions when dealing with text instances that are spatially close enough to each other.  
Such text instances are more common in multi-oriented texts, because most text segments in these texts do not have many spatial changes and have relatively small character and word spaces. Moreover, for the non-Latin multi-oriented texts, there is little or no space between characters in many texts.
In this case, the dense overlapping design of the text segments and our shape-approximation strategy are sufficient to ensure connectivity of text segments. This further demonstrates that the proposed FPNS and SAp strategies are able to boost the performance of bottom-up methods with visual reasoning cues when GCN is removed. However, as shown in Table~\ref{tab:1}, using only visual reasoning (FPNS (GGTR)+SAp) and relational reasoning (GCN+SAp) performs worse than using both visual-relational reasoning cues (FPNS (GGTR)+FPNS (Node)+SAp). The results show that visual-relational reasoning cues produced $0.2\%$-$1.3\%$ performance improvement on several benchmark datasets compared to other reasoning cues, demonstrating that the visual-relational reasoning cues can further boost the performance.

\subsubsection {The Effectiveness on Depicting Long-range Dependency}
To determine the effect of our approach on long-range dependency we performed ablation studies on CTW1500, which has long text instances (see Table~\ref{tab:1}). 
Long-range dependency is achieved mainly by fusing the relational features of GCNs with visual features to generate the GGTR map. Therefore, with/without using FPNS (``GGTR'' in Table~\ref{tab:1}) reflects the two situations of with/without the additional LAT + FD modules to capture long-range dependency. The improvements of F1 obtained on the datasets with line level annotation (CTW1500 and TD-500) also show that %our idea of 
fusing the visual and relational features has contributed to capturing long-range dependency.

Additionally, we compared the quality of the detection results obtained in the Total-Text and CTW1500 datasets with the same settings when the threshold of the IoU was lifted from 0.5 to 0.7. 
As shown in Table~\ref{tab:long text}, applying a more strict matching criterion (\textit{i.e.}, IoU = 0.7) reduced the detection performance on both datasets. 
However, the reduction in the long-text annotated CTW1500 is smaller than that in Total-Text. 
This shows that our method favors long-text instances.

\begin{table}[t]
  \caption{The impact of the annotation types on the proposed method.}
  \label{tab:long text}
  \centering
  \setlength\tabcolsep{0.7pt}
  \begin{tabular}{c|ccc|ccc}
    \toprule
    %\hline
       \multirow{2}{*}{IoU} & \multicolumn{3}{c}{CTW1500 (long-text) } &  \multicolumn{3}{c}{Total-Text} \\  
           %\hline
		\cline{2-7}
		 &P (\%) & R (\%) & F (\%) & P (\%) & R (\%) & F (\%)\\
    \hline
   0.5&89.8&83.3 & 86.4   &84.8 &75.9& 80.1   \\%- &- &-\\
    0.7 &78.7($\downarrow$\tiny11.1)&71.7($\downarrow$\tiny11.6)& 75.0($\downarrow$\tiny11.4)   &69.9($\downarrow$\tiny14.9) &60.0($\downarrow$\tiny15.9)& 64.4($\downarrow$\tiny15.7)  \\%
  \hline
  
\end{tabular}
\end{table}
\begin{table}[!h]
  \caption{The impact of the width of the text segments on the detection accuracy obtained on CTW1500.} %'px' means pixels}
  \label{tab:size}
   \centering
   \setlength\tabcolsep{8pt}
  \begin{tabular}{cccc}
    \toprule
    Width (pixels) &\add{P (\%)} &\add{R (\%)} &F (\%)\\
    \midrule
    1-3 &{\bfseries90.0} & 82.7 &86.2\\
    2-6 &89.8 &{\bfseries83.3} &{\bfseries86.4}\\
    8-12 &89.7 & 83.0 &86.2\\ 
    16-24~\cite{11} &88.3 & 82.3 &85.2\\ 
    $\geqslant$24~\cite{7,8}&87.7 & 82.6 &85.1\\
    \hline
\end{tabular}
\end{table}

\subsubsection {The Impact of the Width of Text Segments}
To assess the impact of the width of text segments on detection accuracy, we conducted an additional ablation study on the CTW1500 dataset, as this dataset contains long curved texts that are sensitive to the setting of text segment width. 

Table~\ref{tab:size} shows that a width between 2-6 pixels achieved the highest F-measure of $86.4\%$. 
It can also be seen that when the width of text segments is larger than 6 pixels, the F-measure decreases. 
We argue that the wider the text segments are, the less well they reflect the characteristics of text, %such as 
\textit{e.g.}, various spaces between text and characters, and that this affects the flexibility of the grouping process, resulting in poorer performance. 
Although a width of 1-3 pixels achieved slightly higher precision of $90.0\%$, 
there are more text segments generated, increasing the burden of NMS. 
Therefore, in our experiments the width of text segments was set to 2-6 pixels.

\begin{table}[t]
  \caption{The Effectiveness of the proposed FPNS mechanism on suppressing false positives and false negatives.}
  \label{tab:fpns}
  \centering
  \begin{tabular}{ccccccc}
    %\toprule
    \toprule
  
		 FPNS&TPs & FPs  & FNs  & \add{P (\%)} & \add{R (\%)} & F (\%)\\
    \hline
    $\times$ &2546& 418 & 522   &85.9 &83.0& 84.4   \\%- &- &-\\
    $\checkmark$  &{\bfseries2559$\uparrow$}&{\bfseries329$\downarrow$} & {\bfseries509$\downarrow$}   &{\bfseries88.6$\uparrow$} &{\bfseries83.4$\uparrow$}& {\bfseries85.9$\uparrow$}  \\%
  \hline
  
\end{tabular}
\end{table}

\subsubsection {The Effectiveness of LAT and FD}

%% WJ
In this work, to realize the proposed visual-relational feature reasoning and capture the long-range dependency between text segments, relational features obtained from GCNs are fused with the visual features obtained from the FPN layers. 
To address the dimensional difference between the relational and visual features, in Section~\ref{fd+lat} we proposed an LAT module to reconstruct the graph convolutional features and a multi-modal Fusion Decoding (FD) module to fuse them, which together generate the GGTR map to regularize text segments. 
To demonstrate the effectiveness of the LAT and FD, we conducted additional experiments on CTW1500. 
The results are shown in Table~\ref{tab:FD}.

Note that, in FPNS (GGTR), the GGTR map is used to rectify false positive/negative text segments, as it is the final visual-relational representation of applying LAT and FD.
The GGTR map is also indirectly involved in the process of FPNS (Node), as it rectifies the text segments before FPNS (Node), leading to the change of relational structure between text segments. Hence, evaluating the two FPNS mechanisms (as shown in the Table~\ref{tab:1}) is equivalent to the evaluation when LAT+FD are in place. FD cannot be applied and tested independently without the LAT module because the LAT process is an essential step for fusing the visual features from FPN and the relational features from GCNs. 
Instead, we tested the effectiveness of FD by replacing the FD module with simple concatenation. 

As shown in  Table~\ref{tab:FD}, 
the FD module can effectively fuse visual and relational features for further FPNS processing. 
The %single 
LAT module with simple concatenation still brings $0.8\%$ gain (improved from $84.4\%$ to $85.2\%$) in F-measure, as the relational information of GCNs is reused in capturing long-range dependency. 
When the FD module replaces %the 
simple concatenation, there is also a further $0.7\%$ gain (improved from $85.2\%$ to $85.9\%$) in F-measure. Compared to simple concatenation, the proposed FD module %can
fuses the visual and relational features more effectively.

\begin{table}[t]
  \caption{The effectiveness of LAT and FD on CTW1500. (w/o: without)} %'px' means pixels}
  \label{tab:FD}
   \centering
  \begin{tabular}{cccc}
    \toprule
    Module &\add{P (\%)} &\add{R (\%)} &\add{F (\%)}\\
    \midrule
    w/o LAT &85.9& 83.0 & 84.4\\
    LAT + Concatenation &87.4& 83.2 & 85.2 \\
    LAT + FD &{\bfseries88.6} &{\bfseries83.4} &{\bfseries85.9}\\
  \hline
\end{tabular}
\end{table}

\subsubsection {The Quantitative and Qualitative Analysis of FPNS Strategies}

To quantitatively and qualitatively validate the effectiveness of the proposed FPNS on suppressing false positives and false negatives, 
ablation studies were conducted on CTW1500, which contains rich samples of long, curved and multi-oriented texts. 
\textbf{i) Quantitative analysis.} 
We first examine the numbers of false positive samples (FPs), false negative samples (FNs) and true positive samples (TPs) in the detection results with and without applying our proposed FPNS strategies. 
As shown in Table~\ref{tab:fpns}, both FPs and FNs have been successfully suppressed to some extent after applying the proposed FPNS strategies, while TPs increase. 
This is also reflected in the increase of the Precision and Recall, and consequently the F-measure. 
Table~\ref{tab:all} shows that after applying the proposed FPNS strategies, both Precision and Recall have surpassed the SOTA. 
\textbf{ ii) Qualitative analysis.} 
Fig.~\ref {1} shows an example where FPs and FNs are suppressed with the proposed FPNS strategies. 
We owe this performance gain to the fact that the proposed FPNS strategies are supervised by the training loss of GCNs, and are less likely to affect true positive samples when suppressing false positives and false negatives.

\add{
Currently, some bottom-up methods~\cite{7,8,11} have put their emphasis on more accurate relational reasoning between text segments, without considering whether those ‘text segments’ are truly text segments or rather interfering backgrounds. 
Even if they attempt to validate text segments, 
whether a text segment is a genuine text segment or not is usually determined by a single modality feature (mostly visual feature). 
When visual feature cannot capture long-range dependency and the `characterness' and `streamline' properties of text, false negatives/positives occur, particularly common in texts with irregular shapes. }

\add{From this perspective, our first FPNS strategy, \textit{i.e.}, `FPNS (GGTR)', jointly considers the relational features from GCNs with the visual feature to rectify false text segments. 
The existing GCN based methods only use relational feature for the purpose of predicting the link between segments, but have ignored that such relational feature is also a ready-made high-dimensional representation of long-range dependency based on the relational connection of text segments. 
During the process of FPNS (GGTR), our newly designed LAT and FD modules reuse these relational features and fuse them with visual features to generate the GGTR map that rectifies the text region.
The relational connection of text segments (acting as the `node' in the graph) can thus reflect the `characterness' and `streamline' properties of text segments.  
Although link prediction is one of the tasks that GCNs are good at and has been well exploited by existing methods, our FPNS (Node) innovatively utilizes the other aspect that GCNs are good at in the arbitrary-shape scene text detection area, \textit{i.e.}, node classification. 
The node classification process in GCNs can be used to further rectify false text segments. This is because the relational feature aggregation between the text segments enable GCNs to rectify a text segment by globally considering the `characterness' and `streamline' of text segments that are in the same relational structure. 
This is an additional cue for determining true positive text segments instead of considering visual feature only. }

\subsubsection {\add{The Impact of the Different Backbones}}
\add{For assessing the impact of the backbone network on the overall performance of our model, we conduct additional studies on CTW1500, Total-Text and ICDAR2015. The results are shown in Table~\ref{tab:bone}. As it shows, when equipped with VGG16, our proposed method has obtained slightly better results than with ResNet50. The same phenomenon has also been reported in~\cite{44,tmm4}. }

\begin{table}[t]
  \caption{\add{Comparison of detection results with different backbones}}
  \label{tab:bone}
  \centering
  \setlength\tabcolsep{1.5pt}
  \begin{tabular}{c|ccc|ccc|ccc}
    \toprule
    %\hline
       \multirow{2}{*}{Backbone} & \multicolumn{3}{c}{CTW1500 } &  \multicolumn{3}{c}{Total-Text}  &  \multicolumn{3}{c}{ICDAR2015}\\  
           %\hline
		\cline{2-10}
		 &P (\%) & R (\%) & F (\%) & P (\%) & R (\%) & F (\%)& P (\%) & R (\%) & F (\%)\\
    \hline
    ResNet50&89.4 &{\bfseries83.5} &86.3    &{\bfseries89.8} &84.8 & 87.2     &{\bfseries92.0} &86.6 &89.2\\%       &- &- &-\\
    VGG16  &{\bfseries89.8} &83.3 & {\bfseries86.4}    &89.4 &{\bfseries85.8}& {\bfseries87.6}    &91.4 &{\bfseries87.7} &{\bfseries89.5}\\%      &83.2 &73.9& 78.3\\
   
  \hline
  
\end{tabular}
\end{table}

\subsubsection {\add{Time Efficiency}} 
\add{Finally, we also report the efficiency of the proposed method. Note that, different scales of the input images and the total number of text segments both affect the running time.  
We collect the statistics of the average processing time per image in each dataset on a workstation with 
a single NVIDIA Quadro P6000 GPU with an Intel(R) Xeon(R) Gold 5122 CPU. 
The running time reported in Table~\ref{tab:time} shows that our model can achieve a decent running time. 
The running time in the curved text detection datasets is faster than those in multi-oriented datasets, since the texts in the latter one are typically smaller than those in the curved text benchmarks. 
As we limit the training to single scale, we have to increase the size of the images to ensure that tiny texts are detected, which increases the inference time to some extent.}

\begin{table}[!h]
  \caption{\add{The time efficiency of the proposed method.}} %'px' means pixels}
  \label{tab:time}
   \centering
   \setlength\tabcolsep{1.5pt}
  \begin{tabular}{cccccc}
    \toprule
    GPU&Datasets & Input Size  & Inference Time  & NMS Time & Total Time\\
    \midrule
    \quad&CTW1500 &640 & 0.28s& 0.51s& 0.79s \\
    Quadro&Total-Text& 640 & 0.22s & 0.58s& 0.80s\\
    P6000&ICDAR2015 & 1280 & 0.68s& 0.89s& 1.57s\\ 
    \quad&MSRA-TD500 & 1280 & 0.67s& 0.90s& 1.57s \\ 
    \quad&MLT2017& 1280 & 0.61s& 0.69s& 1.30s\\
    \bottomrule
\end{tabular}
\end{table}

\section{Limitation}

\begin{figure}[t]  
\subfigure{
\begin{minipage}[t]{0.28\linewidth}
\includegraphics[width=2.7cm,height=2.5cm]{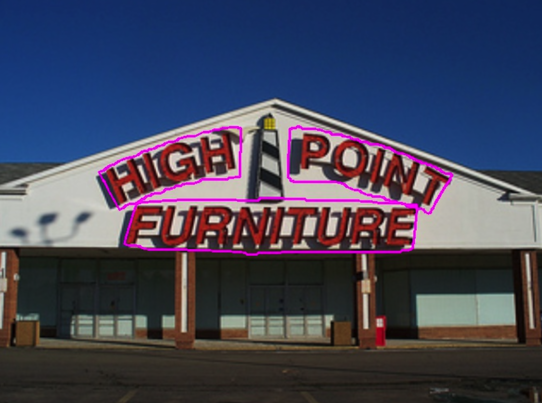}
\label{2aa}
\end{minipage}
} 
\subfigure{
\begin{minipage}[t]{0.28\linewidth}
\includegraphics[width=2.7cm,height=2.5cm]{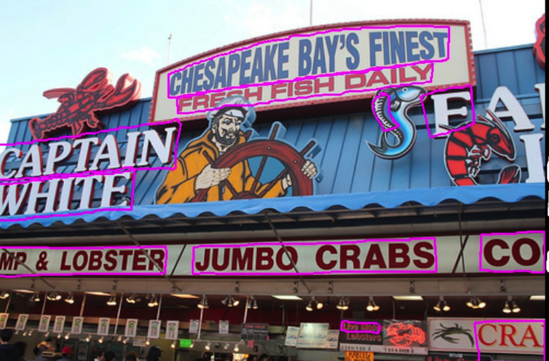}
\label{2aa}
\end{minipage}
}  
\subfigure{  
\begin{minipage}[t]{0.28\linewidth}
\includegraphics[width=2.7cm,height=2.5cm]{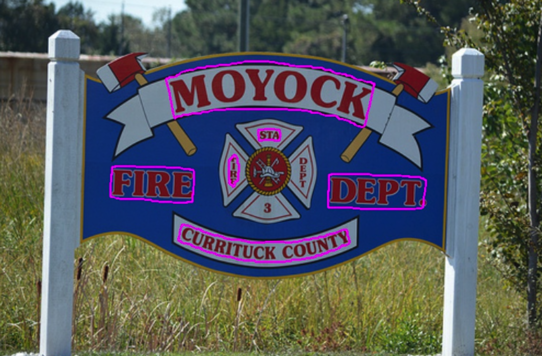}
\label{2b}
\end{minipage}   
}

\vskip -16pt
\subfigure{
\begin{minipage}[t]{0.28\linewidth}

\includegraphics[width=2.7cm,height=2.5cm]{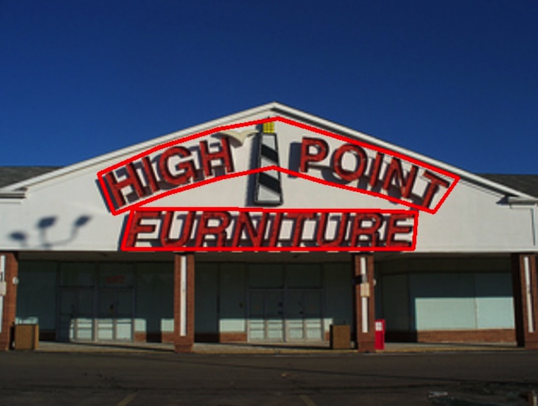}
\label{2aa}
\end{minipage}
}  
\subfigure{  
\begin{minipage}[t]{0.28\linewidth}
\includegraphics[width=2.7cm,height=2.5cm]{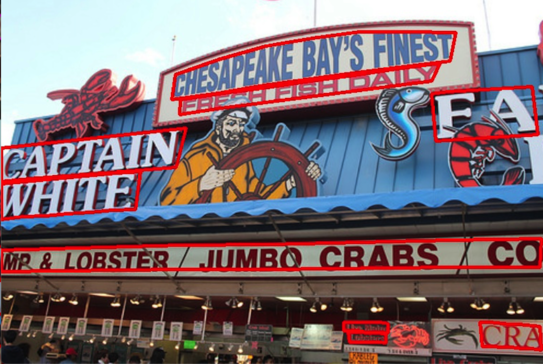}
\label{2b}
\end{minipage}   
} 
\subfigure{
\begin{minipage}[t]{0.28\linewidth}
\includegraphics[width=2.7cm,height=2.5cm]{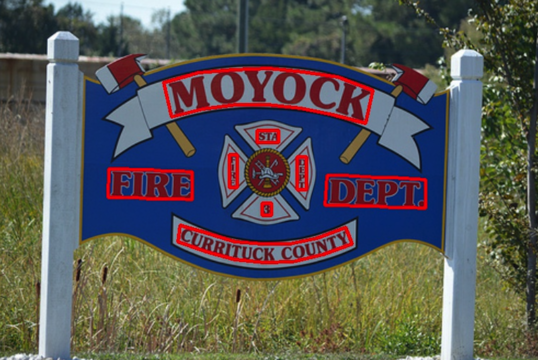}
\label{2aa}
\end{minipage}
}  
%\setlength{\abovecaptionskip}{0.cm}
%\vskip -16pt
\caption{\add{Some failure cases (1st row) and the ground truth (2nd row).} }
\label{fail}
\centering
\end{figure}

\add{Our method can handle texts with large word space thanks to the densely overlapping design of text segments as well as GCNs. However, failure cases happen when long text is separated by non-text objects (see the first and second images in Fig.~\ref{fail}). 
Currently, neither top-down nor bottom-up methods can handle this situation well as the texts are visually separated and how to present the results is debatable even for humans. A potential solution may be to reason according to the semantic information of text. Moreover, failure cases may happen on some text-like objects or super-tiny texts, which are also common challenges for other state-of-the-art methods~\cite{9,42,43}. Examples of such failure cases are shown in Fig.~\ref{fail}.}

\section{Conclusion}

In this paper, aiming to eliminate the error accumulation problem of existing bottom-up arbitrary-shape scene text detection methods, 
we have proposed false positive/negative suppression strategies that take visual-relational feature maps into account to infer grouping of densely designed text segments with regard to GCN's node classification and relational reasoning ability. 
We have also proposed a simple but effective shape-approximation method %module 
to replace the error-prone route-finding process that is currently widely adopted in bottom-up methods. 
The SOTA results obtained on benchmark datasets have demonstrated the effectiveness of our approach on arbitrary-shape text detection, showing 
that bottom-up methods are not necessarily inferior to, but can surpass top-down methods.

% \section*{Acknowledgment}
% This research was supported by National Key R\&D Program of China (2018AAA0100300), UTS FEIT Research Scholarship, China Natural Science Foundation Grant (ID:61976037) and Orbiseed Technology Inc. We thank anonymous reviewers whose comments helped improve and clarify this manuscript.

%\newpage
\bibliographystyle{IEEEtran}
\bibliography{sample-base}

\ifCLASSOPTIONcaptionsoff

\fi

\newpage
\vfill

\end{document}